\documentclass[namedreferences]{SolarPhysics}
\usepackage[optionalrh,solaenum]{spr-sola-addons} 
\usepackage[dvips]{graphicx}
\usepackage{url}                         

\newcommand{\degrees}{\mbox{$^{\circ}$}}		
\newcommand{\textsub}[1]{\mbox{$_{\mbox{\scriptsize #1}}$}}

\newcommand{\eg}{{\it e.g.}}
\newcommand{\ie}{{\it i.e.}}
\newcommand{\vs}{{\it vs.}}

\begin{document}
\begin{article}
\begin{opening}

\title{A Parametric Study of Erupting Flux Rope Rotation}

\subtitle{Modeling the ``Cartwheel CME'' on 9~April~2008}

\author{B. \surname{Kliem}$^{1,2,3}$\sep
        T. \surname{T\"{o}r\"{o}k}$^{4,5}$\sep
        W. T. \surname{Thompson}$^{6}$}

\institute{
 $^{1}$ Insitut f\"{u}r Physik und Astronomie, Universit\"{a}t Potsdam,
        Karl-Liebknecht-Str.\ 24-25, Potsdam 14476, Germany\\
        email: \url{bkliem@uni-potsdam.de}\\
 $^{2}$ Mullard Space Science Laboratory, University College London,
        Holmbury St.~Mary, Dorking, Surrey RH5 6NT, UK\\
 $^{3}$ College of Science, George Mason University, 4400 University
        Drive, Fairfax, VA~22030, USA\\
 $^{4}$ LESIA, Observatoire de Paris, CNRS, UPMC, Universit\'e Paris
        Diderot, 5 place Jules Janssen, 92190~Meudon, France\\
 $^{5}$ now at Predictive Science, Inc., 9990 Mesa Rim Road, Ste.~170,
        San Diego, CA~92121, USA\\
 $^{6}$ Adnet Systems Inc., NASA Goddard Space Flight Center, Code~671,
        Greenbelt, MD~20771, USA}

\begin{abstract}
The rotation of erupting filaments in the solar corona is addressed
through a parametric simulation study of unstable, rotating flux ropes
in bipolar force-free initial equilibrium. The Lorentz force due to
the external shear field component and the relaxation of tension in
the twisted field are the major contributors to the rotation in this
model, while reconnection with the ambient field is of minor
importance, due to the field's simple structure. In the low-beta
corona, the rotation is not guided by the changing
orientation of the vertical field component's polarity inversion line
with height. The model yields strong initial rotations which saturate
in the corona and differ qualitatively from the profile of rotation
\vs\ height obtained in a recent simulation of an eruption without
preexisting flux rope. Both major mechanisms writhe the flux rope axis,
converting part of the initial twist helicity, and produce rotation
profiles which, to a large part, are very similar in a range of
shear-twist combinations. A difference lies in the tendency of twist-driven
rotation to saturate at lower heights than shear-driven rotation. For
parameters characteristic of the source regions of erupting filaments
and coronal mass ejections, the shear field is found to be the
dominant origin of rotations in the corona and to be required if the
rotation reaches angles of order 90 degrees and higher; it dominates
even if the twist exceeds the threshold of the helical kink
instability. The contributions by shear and twist to the total
rotation can be disentangled in the analysis of observations if the
rotation and rise profiles are simultaneously compared with model
calculations. The resulting twist estimate allows one to judge
whether the helical kink instability occurred. This is demonstrated
for the erupting prominence in the ``Cartwheel CME'' on 9~April 2008,
which has shown a rotation of $\approx115\degrees$ up to a height of
$1.5~R_\odot$ above the photosphere. Out of a range of initial
equilibria which include strongly kink-unstable (twist $\Phi=5\pi$),
weakly kink-unstable ($\Phi=3.5\pi$), and kink-stable ($\Phi=2.5\pi$)
configurations, only the evolution of the weakly kink-unstable flux
rope matches the observations in their entirety.

\end{abstract}

\keywords{Corona, Active; Prominences, Dynamics; Coronal Mass Ejections,
Initiation and Propagation; Magnetic fields, Corona; Magnetohydrodynamics}

\end{opening}

\section{Introduction}
\label{s:intro}

The geoeffectiveness of solar coronal mass ejections (CMEs) depends
primarily on two parameters, the velocity and the magnetic orientation
of the CME at the impact on the Earth's magnetosphere. The higher the
CME velocity and the closer its front side magnetic field to a
southward orientation, the more intense the interaction will typically
be. Therefore, understanding the physics that determines these CME
parameters at 1~AU is one of the key issues in space weather research.
This involves the formation and main acceleration of the CME in the
solar corona, as well as its propagation through the interplanetary
space. The particulars of the trigger process also play a role in some
events. It appears that typically the corona is the place where the
basic decisions are made: will the CME be fast or slow, and will it
keep the orientation given by the source, \ie, will its magnetic axis
remain oriented nearly parallel to the photospheric polarity inversion
line (PIL), or will it rotate substantially?

In the present paper we employ the technique of MHD simulation to carry
out a first systematic, but in view of the problem's complexity
necessarily incomplete investigation of a number of processes that cause
and influence changes of CME orientation in the corona. Such changes can
be described as a rotation of the CME volume, more specifically of the
magnetic axis of the flux rope in the CME, about the direction of
ascent. This \emph{rotation} should be distinguished from the possible
rotation of the flux rope about its own axis, referred to as the
\emph{roll effect} \cite{martin:2003,Panasenco&al2011}, which we do not
address here.

Understanding the rotation of erupting flux ropes in the corona is
also relevant for the question which processes trigger the eruptions,
as a substantial rotation may indicate the occurrence of the helical
kink instability (KI); see, \eg, \inlinecite{rust:1996},
\inlinecite{Romano&al2003}, and \inlinecite{rust:2005}. This
instability is one of the candidate mechanisms for the initiation of
CMEs \cite{Sakurai1976,fan:2003,kliem:2004}. It commences when the
twist of the rope exceeds a critical value, 
$\Phi=2{\pi}N>\Phi_\mathrm{cr}$, where $N$ is the winding number of
the field lines about the rope's magnetic axis. The dynamical
evolution of the instability
has shown very good quantitative agreement with a number of well
observed events, which range from confined filament eruptions to the
fastest CME on record \cite{torok:2005,williams:2005}. However,
\inlinecite{isenberg:2007} have pointed out an alternative mechanism
for the rotation of line-tied flux ropes, which relies on the presence
of an external toroidal field component, $B_\mathrm{et}$, due to
sources external to the current in the flux rope and pointing along
the rope, \ie, an external shear field component. The mechanism can
easily be understood in the simplified picture of a current loop in
vacuum field. When the loop legs move out of their equilibrium
position to a more vertical orientation, the cross product of the loop
current with the shear field component yields a sideways Lorentz force
on the legs, which is antisymmetric with respect to
the vertical line that passes through the apex of the loop. This
torque forces the rising top part of the loop to rotate. The effect is
also found in a full fluid description \cite{lynch:2009}. For a given
chirality of the erupting field, it yields the same direction of
rotation as the helical kink. Hence, a comparative study of these two
mechanisms is required before firm conclusions about the occurrence of
the KI can be drawn from observations of flux rope rotations, which is
a further main objective of this paper.

Since the rotations caused by the KI and by the external shear field
point in the same direction, they are difficult to disentangle. In
fact, from a more general perspective, they are of similar nature.
Both cause a writhing of the flux rope which, by conservation of
magnetic helicity, reduces the twist of the rope field lines about the
writhing axis. Consequently, one can expect that observed flux rope
rotations are often consistent with a range of $\Phi$--$B_\mathrm{et}$
parameter combinations which give the writhing of the flux rope by the
helical kink and by the shear field different individual but similar
combined strengths.

Other causes of flux rope rotation include magnetic reconnection with
the ambient field 
\cite{Jacobs&al2009,Shiota&al2010,Cohen&al2010,Thompson2011,Vourlidas&al2011}
and the propagation through the overlying field. The latter comprises
any asymmetric deflection of the rising flux from radial ascent, \eg,
by adjacent coronal holes (see, \eg, \opencite{Panasenco&al2011}), and
the interaction with the heliospheric current sheet
\cite{Yurchyshyn2008,Yurchyshyn&al2009}.

One may conjecture that the generally changing orientation of the PIL
with height in the corona acts similarly to the heliospheric current
sheet at larger heights, \ie, that the upper part of the rising flux
continuously adjusts its orientation to align with the PIL. If this
were the dominant effect, the rotation of erupting flux could be
predicted rather straightforwardly from extrapolation of the
photospheric field, since the overlying field is often close to the
potential field. However, this conjecture is not valid in the lower
corona where $\beta\ll1$, and where the main part of the total
rotation often occurs. We demonstrate this in the Appendix.

The amount of rotation depends on the individual strengths of the four
potentially contributing processes. Each of them is controlled by more
than a single parameter. This is immediately obvious for the torque by
the shear field, which must depend on the height \emph{profile}
$B_\mathrm{et}(z)$, and for the reconnection, which is sensitive to
the \emph{structure} of the ambient field, \ie, whether the field is
bipolar, quadrupolar, or multipolar and whether the orientation of the
line between the resulting new footpoints of the erupting flux differs
strongly from the original orientation. The rotation by the KI does
not only depend on the initial flux rope twist,
$\Phi-\Phi_\mathrm{cr}$, but also on the strength and height profile
of the overlying field \cite{Torok&al2010}. If the overlying field
decreases only slowly with height, then the upward expansion develops
slowly and, accordingly, its contribution to the relaxation of the
field line tension is initially weak. The relaxation is then primarily
accomplished by a strong rotation at small heights. In the opposite
case of very strong upward expansion, the rotation is distributed
across a large height range, which also increases the likelihood of
further changes by the onset of reconnection (see
\opencite{Lugaz&al2011} for an example). The effect of the
heliospheric current sheet can be expected to depend on the angle with
the top section of the flux rope's axis, on the horizontal elongation
of the CME (whether its horizontal cross section is very elliptical or
more nearly circular), and on the magnetic pressure of the CME
relative to the pressure of the interplanetary plasma.

Moreover, the total rotation experienced by an erupting flux rope
likely depends also on the dynamics of its evolution. For example, a
torque strongly localized at low heights, operating on a still small
loop, may hurl the flux around more efficiently than a torque which is
distributed across a large height range. As another example, in a
complex (multipolar) coronal environment the sequence and strength of
reconnection with the ambient field may strongly depend upon the
height profiles of the rope's angular and rise velocities caused by
other processes, \eg, by an ideal MHD instability. The relative
velocity between reconnecting flux systems controls how strongly the
reconnection with the ambient field is driven. Hence, quantitative
studies of flux rope rotation face a very high degree of complexity.

Here we focus on two mechanisms that can cause strong rotations in the
corona, the helical kink instability and the torque exerted by an
external shear field component. By comparing a parametric study of
both mechanisms in a force-free, line-tied flux rope equilibrium with
the data of a well observed, strong\-ly rotating erupting prominence,
we demonstrate that their contributions can be disentangled to some
degree. We also demonstrate the very strong influence of the ambient
potential field's height profile on the amount of rotation by the KI,
and briefly address the influence of reconnection between the CME flux
rope and the ambient field on the rotation.

\begin{figure}
\begin{center}
\includegraphics[width=0.9\textwidth]{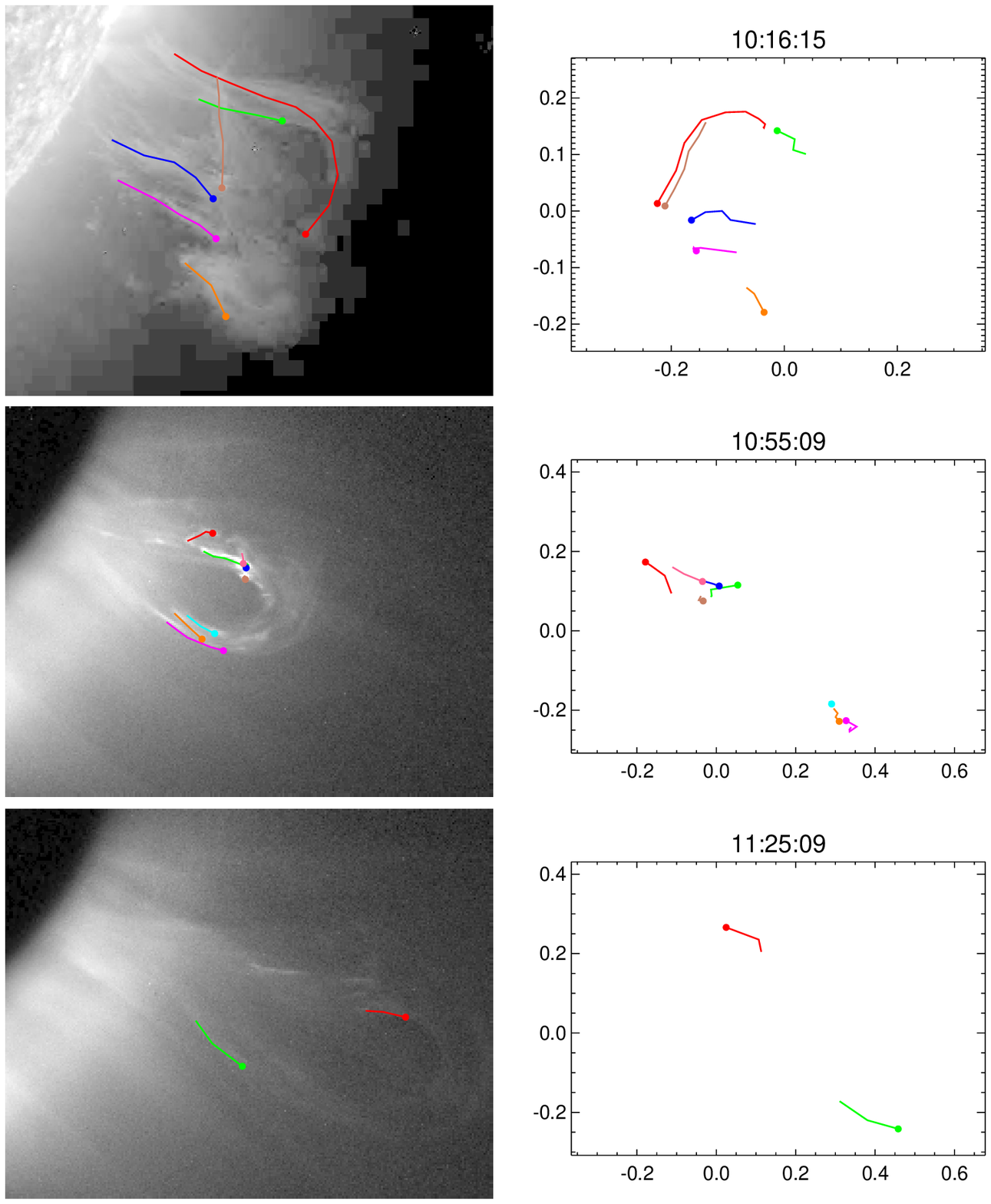}
\end{center}
\caption[]
{Images and plots of the prominence eruption at 10:16~UT, as seen by
 the EUVI-Ahead telescope in the 304~{\AA} channel, and at 10:55 and
 11:25~UT, as seen in white light by COR1-Ahead. The prominence apex
 has reached heights of 0.56, 1.6, and $2.3~R_\odot$ above the
 photosphere at these times. The right panels display the
 reconstructed three-dimensional position of the marked prominence
 threads, using a reprojection to a viewpoint at the position of
 radial CME propagation, Stonyhurst longitude $98\degrees$ west
 (relative to Earth) and latitude $24\degrees$ south, where the
 counterclockwise rotation is apparent. The axes are in units of solar
 radii.}
\label{f:obs}
\end{figure}

This investigation was stimulated by the analysis of the strong
rotation in a prominence eruption and CME on 9~April 2008,
occasionally referred to as the ``Cartwheel CME'', in
\citeauthor{Thompson&al2011} (\citeyear{Thompson&al2011}, in the
following Paper~I). Their stereoscopic reconstruction revealed the
height-rotation profile of the erupting filament/prominence in the
core of a CME for the first time \cite{Thompson&al2009}.
This profile provides a strong constraint for the numerical 
modeling. In combination with the further observations of the event,
it allows us to infer the major causes of the rotation and the range
of source parameters compatible with the data. The analysis of Paper~I
has given the following results relevant for the present study. The
prominence erupted from the remnants of NOAA active region (AR) 10989
close to the west limb and appeared as a flux rope -- a single, weakly
to moderately twisted loop -- throughout the height range covered by
the STEREO EUVI and COR1 telescopes \cite{howard:2008}, \ie, up to
$4~R_\odot$ from Sun center. It rotated counterclockwise by
$\approx\!115\degrees$ up to a heliocentric height of $2.5~R_\odot$,
where the rotation leveled off. Two thirds of this rotation were
acquired within $0.5~R_\odot$ from the photosphere. The data indicate
a subsequent gentle backward rotation by $\approx\!15\degrees$ in the
height range up to $3.3~R_\odot$. In addition, the analysis of STEREO
COR2 data in \inlinecite{Patsourakos&Vourlidas2011} demonstrated that
a flux rope structure is consistent also with the three-dimensional
shape of the CME at a heliocentric distance of $13~R_\odot$, where it
had changed its orientation by a total of $150\degrees\pm7\degrees$
from the original one, most likely by further counterclockwise
rotation. At this stage the erupting flux was very closely aligned
with the heliospheric current sheet above the active region. The
prominence was initially accelerated mainly horizontally along the
filament channel. This gradually turned into a radial propagation at a
position $\approx$\,98W24S as seen from Earth,
$15\degrees\mbox{--}20\degrees$ away from the original location. The
prominence experienced most of its upward acceleration in the
heliocentric height range up to $\sim\!2.5~R_\odot$ and reached a
velocity of $\sim\!400$~km\,s$^{-1}$ in the COR2 field of view. At the
same time, the leading edge of the CME accelerated to over
700~km\,s$^{-1}$ \cite{Landi&al2010}. Representative images of the
prominence from STEREO \emph{Ahead}, which had the best perspective at
the structure, and the corresponding three-dimensional reconstructions
of the location of several prominence threads are compiled in
Figure~\ref{f:obs} (from Paper~I). The rotation (height-rotation)
profile and the rise (time-height) profile are included below in the
observation-simulation comparisons (Figures~\ref{f:rotation} and
\ref{f:rise}, respectively).

As already noted above, we focus our attention here on the coronal
evolution of this event, leaving the interaction with the heliospheric
current sheet for future investigation. Moreover, we exclude the
possible slight backward rotation by $\approx15\degrees$ in the COR1
height range from our modeling, since we are interested in the
generally important effects which cause significant rotations in the
corona. This part of the rotation is not fully certain, and, if real,
it was likely caused by the particular structure of the large-scale
coronal field above the active region, which nearly reversed its
horizontal direction at heights $\gtrsim\!0.3~R_\odot$ above the
photosphere (Paper~I). Thus, we will consider a saturation of the
modeled rotation at angles near 115\degrees\ and heights
$h\approx(1.5\mbox{--}2.3)~R_\odot$ above the photosphere to be in
agreement with the observation data. Furthermore, we will disregard
the initial nearly horizontal motion of the prominence along the PIL.

The combined effects of flux dispersal and foreshortening in the
course of the source region's rotation to the solar limb made it
impossible to obtain a well-defined estimate of the distance between
the main flux concentrations in the bipolar region at the time of the
eruption, which is a parameter of strong influence on the height
profile of the ambient potential field. Only a relatively wide range
of $\sim\!(40\mbox{--}150)$~Mm could be estimated by extrapolating 
the region's evolution in the course of its disk passage through the
final three days before the event. It will be seen that this range
still sets a useful constraint on the modeling.

In the following we model the radial propagation of the prominence in
the Cartwheel CME in the coronal range of heights as the MHD evolution
of an unstable force-free and line-tied flux rope
(Section~\ref{s:model}). A parametric study of the resulting rotation
and rise, focusing on the rotation caused by the helical kink
instability and by the external shear field, is compared with the
observation data, to constrain the parameters in the source of the
event and to study whether the relative importance of these mechanisms
can be disentangled and individually estimated
(Section~\ref{s:comparison}). The discussion in
Section~\ref{s:discussion} addresses the simplifying assumptions made
in the modeling and differences to earlier relevant work.
Section~\ref{s:conclusions} gives our conclusions. The Appendix
relates the rotation of erupting flux ropes in low-beta plasma to the
changing orientation of the PIL with height.


\section{Numerical Model}
\label{s:model}

\begin{figure}   
\centering
\includegraphics[width=0.9\textwidth,clip=]{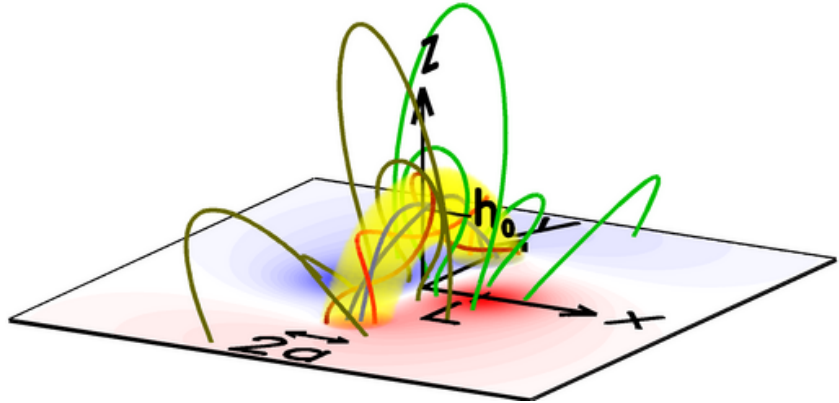}
\caption[]
{Visualization of the modified Titov-D\'emoulin flux rope equilibrium
 used as the initial condition in the simulation runs of this paper;
 here with an average twist $\Phi=3.5\pi$ as in Figure~\ref{f:3.5pi}.
 The current channel is rendered as a yellow transparent volume. Blue
 field lines run near the magnetic axis of the flux rope (where the
 local twist is $2\pi$), red field lines lie in the flux surface at a
 distance to the axis where the local twist equals the average twist.
 Green and olive field lines show the ambient potential field.
 Contours of the magnetogram, $B_z(x,y,0)$, are shown in the bottom
 plane. The torus of major radius $R$ and minor radius $a$ is
 submerged by a distance $d$, resulting in the apex height $h_0=R-d$
 and the distance of each flux rope footpoint from the origin
 $D_\mathrm{f}=(R^2-d^2)^{1/2}$. A bipole, whose components are
 located at $({\pm}L,0,-d)$, is the source of the external poloidal
 field component $B_\mathrm{ep}$; see Figure~2 in
 \inlinecite{titov:1999} for its visualization. A pair of
 antiparallel, vertically oriented dipoles, placed under the
 footpoints of the flux rope at $(0,{\pm}D_\mathrm{f},-5h_0)$,
 provides the source of the external toroidal (shear) field component
 $B_\mathrm{et}$.}
\label{f:TD99_3.5pi}
\end{figure}

We carry out a series of MHD simulations similar to the CME simulation
in \inlinecite{torok:2005}. The prominence is modeled as a section of
an approximately force-free toroidal current channel embedded in
external current-free (potential) field, which represents a
modification of the approximate force-free equilibrium by
\inlinecite{titov:1999}.  The current channel creates a flux rope
structure of the magnetic field which has a somewhat larger cross
section than the channel and is enclosed by a quasi-separatrix layer
in the interface to the surrounding field of arcade structure.  The
chirality of the flux rope is chosen to be left handed, so that the
rotation will be counterclockwise \cite{green:2007}. The poloidal
component of the external field, $B\textsub{ep}$, is due to a pair of
subphotospheric magnetic point sources, which produce a pair of flux
concentrations (``sunspots'') to the sides of the flux rope (the
``prominence'') in the magnetogram. This field component holds the
current channel in equilibrium; its strength at the position of the
rope is exactly proportional to the current  in the rope.
Consequently, only its spatial profile, determined by the spacing
between its sources, can be freely varied. The toroidal component of
the external field, $B\textsub{et}$, is due to a pair of
subphotospheric dipoles, positioned under the footpoints of the flux
rope such that the field lines of $B\textsub{et}$ run parallel to the
magnetic axis of the rope to a very good approximation.  Therefore,
$B\textsub{et}$ introduces only very minor Lorentz forces in the
initial configuration, which quickly decrease by numerical relaxation
at the beginning of each run, so that the strength of $B\textsub{et}$
can be chosen freely within a wide range. We will refer to the
external toroidal field also as the shear field component. Here it
decreases faster with height than the external toroidal field in the
original Titov-D\'emoulin equilibrium. A visualization of the
configuration is shown in Figure~\ref{f:TD99_3.5pi}.

We integrate the ideal MHD equations but neglect pressure, as
appropriate in the active-region corona, and gravity, because the
hydrostatic pressure profile along the field is not essential for the
flux rope rotation, which is driven by the Lorentz force. These
simplifications yield maximum freedom in the scalability of the
simulation results to the data. Magnetic reconnection can occur due to
the numerical diffusion of the field in regions of strong gradients. 
The initial density is specified as
$\rho_0({\bf x})=|{\bf B}_0({\bf x})|^{3/2}$, where
${\bf B}_0({\bf x})$ is the initial magnetic field. This yields a slow
decrease of the Alfv\'{e}n velocity with height, as in the corona. The
box is a cube 64~units long on each side, significantly larger than in
our previous simulations and in each direction at least twice as large
as the biggest size of the structures that will be compared to the
data. It is resolved by a nonuniform, fixed Cartesian grid with a
resolution of 0.04~units in the central part of the box (a factor of 2
coarser than in \citeauthor{torok:2005}, 2005). Rigid boundary
conditions are implemented at the top and side boundaries, while very
small velocities are permitted in the bottom boundary. Initially the
torus lies in the plane $\{x=0\}$. The MHD variables are normalized by
the initial apex height of the flux rope axis, $h_0$, by the initial
field strength $B_0$, density $\rho_0$, and Alfv\'{e}n velocity
$V_\mathrm{A0}$ at this point, and by the corresponding quantities
derived thereof, \eg, the Alfv\'{e}n time
$\tau_A = h_0/V_\mathrm{A0}$. Thus, the initial apex height of the
axis of the current channel and flux rope serves as the length unit.

The parameters of the initial configuration are largely chosen as in
\inlinecite{torok:2005}. We fix the major radius of the torus at
$R=1.83$, the depth of the torus center at $d=0.83$ and the
pre-normalization strength of the point sources at $q=10^{14}$~Tm$^2$
in all runs. For a base set of the simulation series, discussed below
in Figures~\ref{f:5pi}--\ref{f:rotation} and
\ref{f:rise}--\ref{f:comp}, we further fix the distance of the point
sources from the $z$ axis at $L=0.83$ (in units normalized such that
$h_0$ is unity). This value lies in the middle
of the estimated range for the corresponding distance of the flux
concentrations in AR~10989, given above, when the scaling
$h_0=0.077~R_\odot$ adopted in Section~\ref{ss:rotation_profile} is
applied. It also agrees with the settings in several previous
investigations (\eg, \opencite{torok:2005}; \opencite{Torok&al2010}),
facilitating comparisons. Variations of this parameter will be
considered in the range $L=0.42\mbox{--}2.5$. We vary the minor radius
of the toroidal current channel, $a=0.32\mbox{--}0.62$, and the
strength of the external toroidal field,
$B_\mathrm{et}/B_\mathrm{ep}=0\mbox{--}1.06$ at the flux rope apex
${\bf x} = (0,0,1)$, to obtain a range of values for the average twist
of the current channel, $\Phi=(2.5\mbox{--}5.0)\pi$, and for the
strength of the shear field component. The twist is influenced by both
$a$ and $B_\mathrm{et}$, with $a$ having the stronger influence within
the considered range of parameters. The twist values quoted in this
paper represent the initial twist averaged over the cross section of
the current channel in the manner described in
\inlinecite{torok:2004}.

The range of the initial average twist is chosen such that unstable
and stable configurations with respect to the helical kink mode are
included. The first group is unstable from the beginning of the
simulation. Nevertheless, a small upward initial velocity perturbation
is applied in the vicinity of the flux rope apex (typically ramped up
to $0.05~V_\mathrm{A0}$ over $5~\tau_A$ and then switched off), to
ensure that the instability displaces the apex upwards, \ie, downward
kinking is excluded in these runs which are intended to model CMEs.

For the geometric parameters of the system specified above, the flux
rope is initially stable with respect to the torus instability
\cite{kliem:2006,torok:2007}.  However, the helical kink instability
lifts the rope into the torus-unstable range of heights
($h\gtrsim2~h_0$), from where the torus instability accelerates its top
part further upwards.\footnote{The torus instability can be considered
as a lateral kink of the current channel. However, we choose ``kink''
and ``KI'' to refer exclusively to the helical kink mode in this
paper.} The kink-stable cases require that the upward velocity
perturbation is applied for a longer time, lifting the apex into the
torus-unstable range. This allows us to study the influence of the
shear field on the rotation in the absence of the helical kink
instability, using uniform geometrical parameters of the initial flux
rope (except for the minor flux rope radius $a$) in all runs. An
initial velocity perturbation very close to the required minimum value
is applied in each of these cases, to ensure nearly uniform conditions
at the onset of the instability throughout the series. The values at
the end of the ramp phase stay below $0.12~V_\mathrm{A0}$ for all runs.
The flux rope velocity falls back to a much smaller value (typically
$\approx\!0.01~V_\mathrm{A0}$) immediately after the perturbation is
switched off. The growing instabilities then accelerate the apex to
peak upward velocities in the range
$\max\{u_\mathrm{a}\}\approx(0.4\mbox{--}0.7)~V_\mathrm{A0}$, far
higher than the initial perturbation.

On the Sun, the initial lifting of the flux can occur by a variety of
effects in addition to the helical kink mode, as has been demonstrated
by numerical simulations. These include the shearing and twisting of
the coronal field by photospheric flows (\eg,
\opencite{Mikic&Linker1994}; \opencite{torok:2003}), reconnection
associated with flux cancellation in the photosphere (\eg,
\opencite{Aulanier&al2010}; \opencite{Amari&al2010}), and reconnection
with newly emerging flux \cite{Chen&Shibata2000}.

The observations of the Cartwheel event indicate a gradual doubling of
the prominence height prior to the eruption (Paper~I). The initial
lifting of the flux rope apex in the simulations due to the applied
perturbation is much smaller for all kink-unstable runs and stays in
the range up to this value for the kink-stable cases, except for the
run with the highest shear field ($\Phi=2.5\pi$,
$B_\mathrm{et}/B_\mathrm{ep}=1.06$), which requires a lifting to
$2.6~h_0$.

\section{Comparison Simulations-Observations}
\label{s:comparison}

\subsection{Dependence of Flux Rope Rotation on Twist and Shear}
\label{ss:rotation_profile}

\begin{figure}   
\centering
\includegraphics[width=0.63\textwidth]{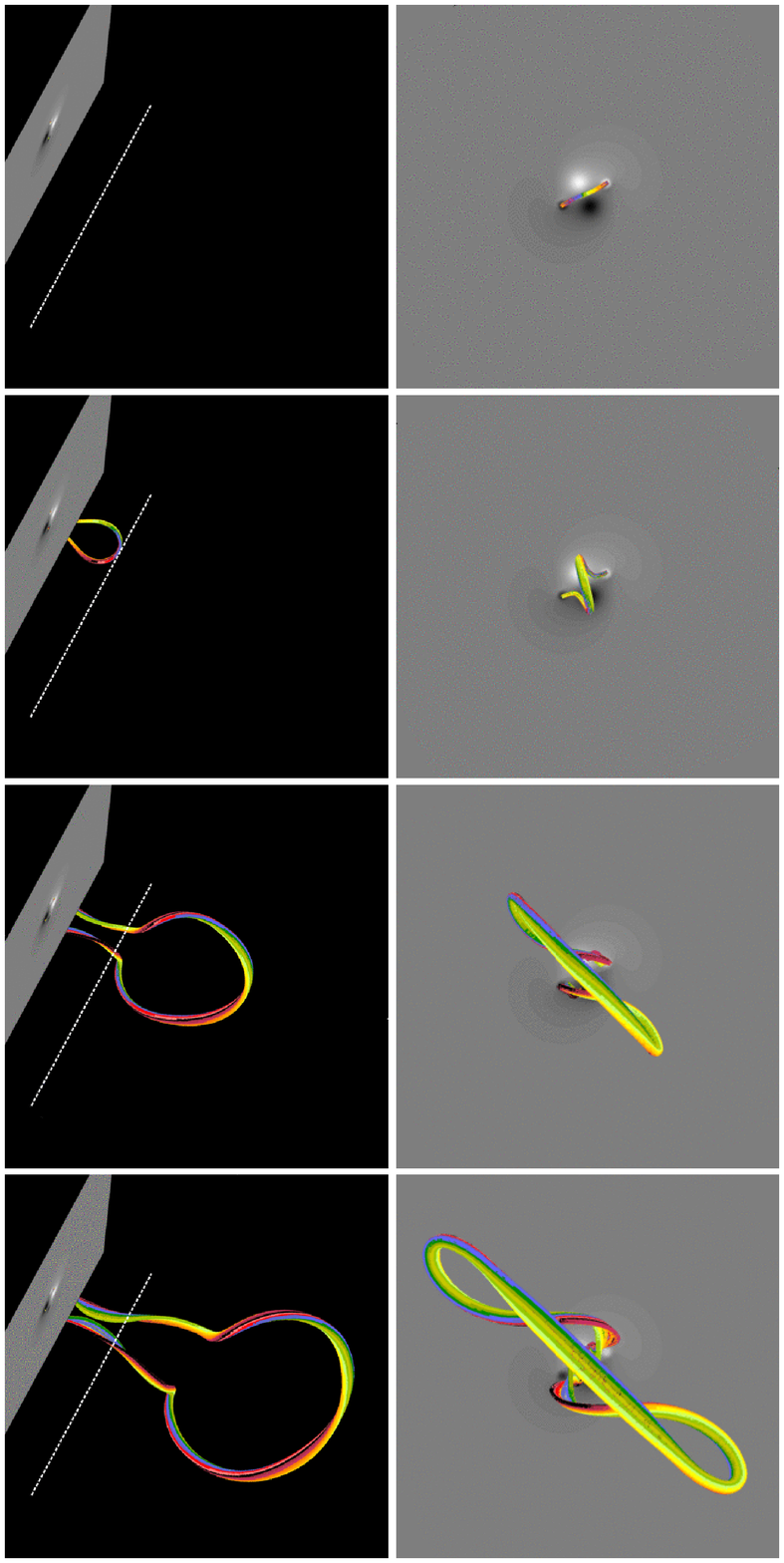}
\caption[]
{Snapshots of an erupting and rotating, strongly kink-unstable flux
 rope. The initial average twist is $\Phi=5\pi$ and the shear field
 component at the initial flux rope apex position is given by
 $B_\mathrm{et}/B_\mathrm{ep}=0.42$. Field lines in the core of the
 rope, traced downward from the apex, are shown in the height range
 $0\le z\lesssim30$, using the same two perspectives as for the
 observation data in Figure~\ref{f:obs} (in the left panels the line
 of sight makes an angle of 26{\degrees} with the $y$ axis, and the
 $z$ axis is tilted away from the observer by 8{\degrees}, while the
 right panels present a vertical view with an initial angle between
 the flux rope axis and the east-west direction of 26{\degrees}).
 The magnetogram, $B_z(x,y,0,t)$, is displayed in grayscale (seen from
 below in the left panels). The dotted line indicates where
 the edge of the COR1 occulting disk is located if the distance
 between the flux rope footpoints in the simulation,
 $2D_\mathrm{f}=3.3~h_0$,
 is scaled to the value of 175~Mm estimated in Paper~I. Using this
 scaling, the simulated heights of $h=1$, 7.3, 21, and $30~h_0$ (at
 $t=0$, 36, 64, and $84~\tau_A$) translate to heights of 0.077, 0.56,
 1.6, and $2.3~R_\odot$ above the photosphere, reached at 10:16, 10:55,
 and 11:25~UT (for rows 2--4), respectively.}
\label{f:5pi}
\end{figure}

\begin{figure}   
\centering
\includegraphics[width=0.63\textwidth]{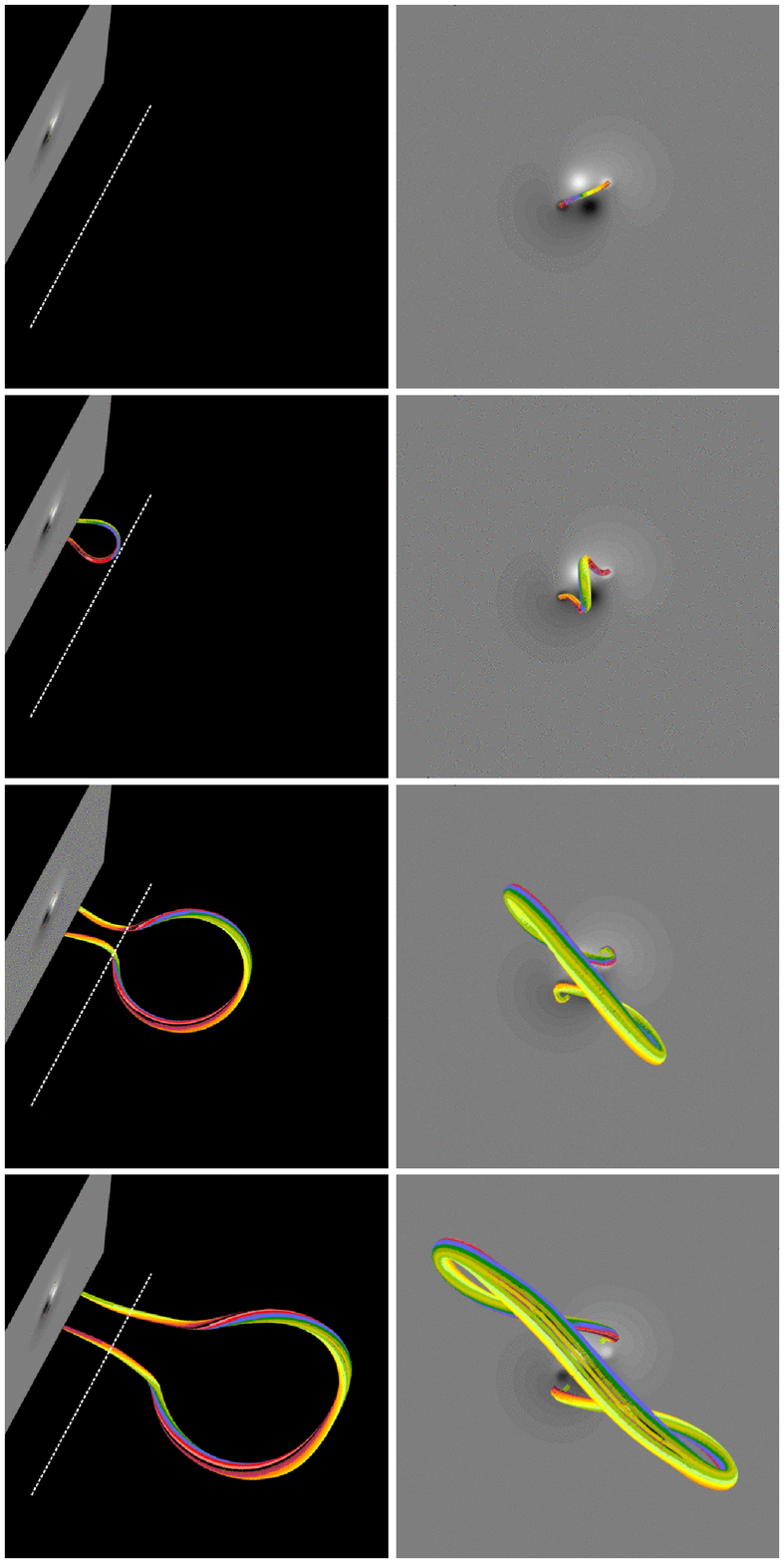}
\caption[]
{Same as Figure~\ref{f:5pi} for a weakly kink-unstable case with
 initial avarage twist $\Phi=3.5\pi$ and shear field
 $B_\mathrm{et}/B_\mathrm{ep}=0.67$. The flux rope is shown at the
 simulation times $t=0$, 50, 80, and $97~\tau_A$ which yield the same
 heights as the snapshots in Figure~\ref{f:5pi}, corresponding to the
 same observation times.}
\label{f:3.5pi}
\end{figure}

We begin with a case that involves a clear helical kink instability,
as one would expect at first sight from the considerable rotation
observed in the Cartwheel event. The initial average twist is chosen
to be $\Phi=5\pi$, a value used previously in the successful modeling
of several filament/prominence eruptions
\cite{williams:2005,torok:2005}. Even with this considerable amount of
twist (and with the sunspot semi-distance $L=0.83$), we find that a shear
field is required to reach the observed rotation. Figure~\ref{f:5pi}
shows the resulting rotation of the flux rope, which reaches the
observed value of $115\degrees$ and is a combined effect of the
helical kink instability and the shear field. The field lines
visualize a flux bundle in the core of the rope which runs slightly
($\approx\!5\%$) under the rope axis in its top part. This is a likely
location for prominence material within a flux rope. Moreover, this is
the only selection that allows a favorable comparison with the
observed flux rope shape for the weakly twisted case shown below in
Figure~\ref{f:2.5pi}, while the more strongly twisted cases are less
sensitive to this vertical offset. Therefore, we adopt this selection
as a uniform choice for Figures~\ref{f:5pi}-\ref{f:2.5pi} which
compare the flux rope rotation for different twist values. The field
lines are displayed from perspectives identical to the STEREO images
and reconstructions in Figure~\ref{f:obs}.

Two characteristic morphological features apparent in the COR1 data in
Figure~\ref{f:obs} are weakly indicated in the simulation: the initial
teardrop-like appearance and the elongated shape at large heights
(relatively narrow in the horizontal direction). The right panels show
that the teardrop shape is a projection effect. The legs of the
erupting rope approach each other near the edge of the occulting disk
only in projection; they are displaced along the line of sight and
actually moving away from each other. The elongated shape is largely
also due to the strong rotation.

The legs of the rope appear ``wiggly'', which results from two
effects. First, they reconnect with the ambient field in the vertical
current sheet under the flux rope apex in the interval
$t\approx(32\mbox{--}65)~\tau_A$, which corresponds to apex heights
$h\approx(5\mbox{--}21)~h_0$; with the reconnection proceeding at much
lower heights inside the edge of the COR1 occulting disk. This leads
to a bend in the reconnected flux rope: the field lines have
relatively small curvature within the legs of the expanded original
rope above the reconnection point but run along a more helical path in
the ambient field just outside the original rope below the
reconnection point. This bend and the more helical shape of the field
lines below it relax upward, along with the overall upward expansion
of the reconnected flux rope. Since the flux rope apex has reached a
considerable upward velocity, $u_\mathrm{a}\lesssim0.5~V_A$, the bend
needs a large height range for its propagation to the top of the rope.
It is located slightly above the dotted line in the third snapshot
pair of Figure~\ref{f:5pi} and at $h\gtrsim15~h_0$ in the final
snapshot pair. The plots on the right hand side show that the new
footpoints of the rope are displaced in counterclockwise direction
from the original ones, thus contributing to the overall
counterclockwise rotation of the rope. However, this contribution is
only a minor one; the major part of the total rotation occurs before
the flux rope legs reconnect (which can be seen by comparison with
Figure~\ref{f:rotation} below). This reconnection is similar to the
second and third reconnections described in
\citeauthor{Gibson&Fan2008} (\citeyear{Gibson&Fan2008}, their
Section~4.1) and will be addressed in more detail in a future
investigation. Second, at the given relatively high value of the
twist, the dominant wavelength of the helical kink mode is
considerably shorter than the flux rope, so that the characteristic
helical shape develops clearly.

\begin{figure}   
\centering
\includegraphics[width=0.63\textwidth]{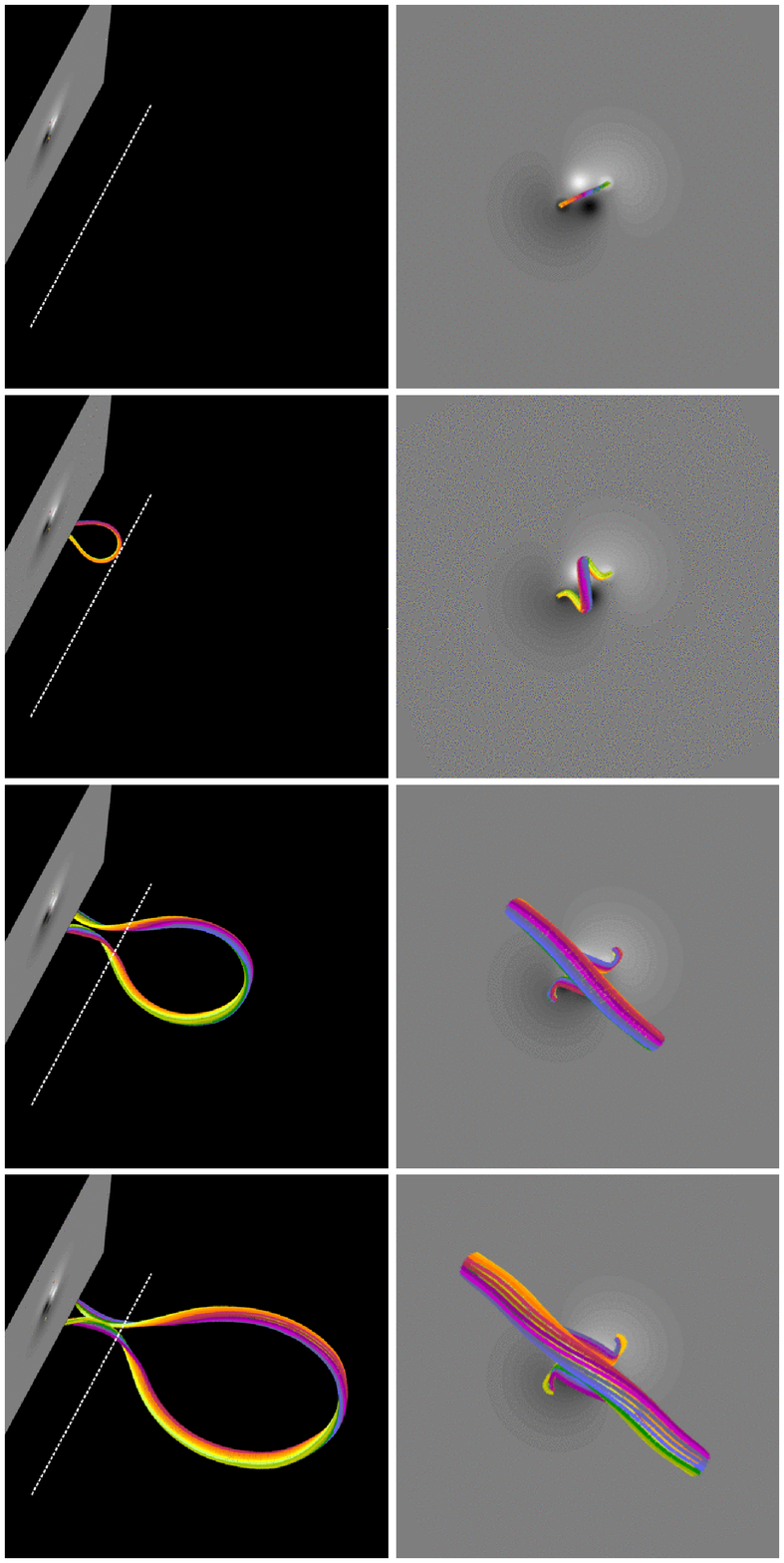}
\caption[]
{Same as Figure~\ref{f:5pi} for a kink-stable case with initial
 avarage twist $\Phi=2.5\pi$ and shear field
 $B_\mathrm{et}/B_\mathrm{ep}=1.06$. The flux rope is shown at the
 simulation times $t=0$, 77, 109, and $128~\tau_A$ which yield the same
 heights as the snapshots in Figure~\ref{f:5pi}, corresponding to the
 same observation times.}
\label{f:2.5pi}
\end{figure}

Figure~\ref{f:3.5pi} shows the evolution in a second run where the KI
develops only weakly, using a moderate, only slightly supercritical
value of the initial average twist, $\Phi=3.5\pi$. A stronger shear
field is chosen, so that the same total rotation is achieved. The
overall properties -- accelerated rise into an ejection (CME) and very
strong rotation -- are identical to the run shown in
Figure~\ref{f:5pi}. The morphological details, such as the teardrop
shape at small heights and the elongated shape at large heights, match
the data slightly better. The indications of wiggly shape at large
heights remain  weak. Reconnection of the flux rope legs with the
ambient field occurs here as well, but the resulting changes in the
shape of the flux rope are weaker, since not only the field lines in
the rope are less twisted but also the ambient field is less helical,
due to the larger $B_\mathrm{et}$. This morphological difference to
the strongly twisted flux rope is one aspect that may allow to
distinguish between rotations with strong and weak involvement of the
helical kink in observed events. The field line shapes in the present
case conform slightly better to the inclination of the prominence
threads with respect to the axis of the flux rope in the COR1 data in
Figure~\ref{f:obs}, but this difference is not sufficiently clear to
be decisive by itself. Moreover, it depends to a considerable degree
upon which part of the erupting flux was outlined by prominence
material in the considered event and on the selection of field lines
in the plots.

Figure~\ref{f:2.5pi} presents a case with subcritical flux rope twist,
$\Phi=2.5\pi$, where the kink instability cannot develop and an even
stronger shear field is needed to achieve a similar rotation. Here the
parameters were chosen such that the rotation matches the observations
as well as the other two runs in the height range $h\lesssim20~h_0$,
with the total rotation of the rope's magnetic axis at $h=30~h_0$
exceeding the rotation in those runs by 20--25~degrees. The elongated
teardrop shape at intermediate and large heights yields the best match
of the three runs shown in Figures~\ref{f:5pi}--\ref{f:2.5pi}.
However, this is only the case because a flux bundle slightly under
the magnetic axis of the flux rope is selected in the visualization.
If instead a set of field lines encircling the flux rope axis is
chosen, then the high total rotation at the apex height $h=30~h_0$
leads to an inverse teardrop shape (narrow at the apex, because at
this point the view is nearly along the axis of the rotated flux
rope), which is inconsistent with the observations. Again, since it is
not known which parts of the erupting flux (rope) were filled with
prominence material in the event to be modeled, these morphological
comparisons, by themselves, do not allow to rule out the kink-stable
run shown in Figure~\ref{f:2.5pi}.

The similar total rotations in the three simulations confirm that both
twist and shear belong to the key parameters which determine the
amount of rotation in erupting flux ropes. To analyze this further, we
consider a set of characteristic cases from our series of simulation
runs with varying strength of the two effects. For each of the twist
values $\Phi=5.0$, 3.5, and $2.5\pi$, we vary the shear field
$B_\mathrm{et}$ from the respective best fitting value used in
Figures~\ref{f:5pi}--\ref{f:2.5pi}. All runs use the same sunspot
semi-distance $L=0.83$ and, hence, the same external poloidal field
$B_\mathrm{ep}$. The variation of $L$ will be considered in
Section~\ref{ss:poloidal_field}.

The rotation of the flux rope in the simulations is measured in two
ways. At low heights it is taken from the changing orientation of the
magnetic axis at the apex of the flux rope. As the flux rope rises,
the apex orientation oscillates increasingly, due to the upward
propagation of Alfv{\'e}nic perturbations which result from the
dynamic onset of reconnection in the vertical current sheet under the
rope (the relaxation of the bend in the reconnected field lines
mentioned above). The right panels at the two final heights in
Figures~\ref{f:3.5pi}--\ref{f:2.5pi} indicate the resulting
oscillations of the field orientation at the apex with respect to the
bulk orientation of the flux rope's upper part. Therefore, at larger
heights we simply use the direction of the horizontal line connecting
the flux rope legs at the height where they are most distant from each
other. This measurement filters away most of the oscillating
variations, which are also not captured by the observed rotation
data derived in Paper~I and replotted in Figure~\ref{f:rotation}. The
difference between the two measurements remains less than 5~percent in
a height range $\Delta h\sim(3\mbox{--}6)~h_0$ around $h\sim10~h_0$,
except for the most strongly rotating and oscillating case in the
series ($\Phi=5\pi$, $B_\mathrm{et}/B_\mathrm{ep}=0.63$) where it
reaches $\approx\!10$~percent. Linear interpolation between the two
measurements for each simulation run is applied in the appropriate
range of small difference to match them smoothly. (The method to
estimate the rotation angle at large heights fails for one of the runs
in Figure~\ref{f:rotation} ($\Phi=2.5\pi$, $B_\mathrm{et}=0$), where
reconnection of the flux rope legs with the ambient field leads to
jumps that are larger than the oscillations of the magnetic axis at
the apex. For this run, whose rotation profile differs strongly from
the observed one, we include the rotation angle only at low heights,
to show the trend.)

In order to compare the simulated rotation profiles with the
observations, a scaling of the length unit in the simulations to
distances on the Sun is required. For this purpose, we set the
distance between the footpoints of the flux rope in the simulation,
$2D_\mathrm{f}=3.3~h_0$, equal to the estimated length of the flux which
holds the prominence, 175~Mm (Paper~I). This is independent of the actual
prominence shape. The apex height of the toroidal Titov-D\'emoulin
flux rope, our length unit, tends to be somewhat high in comparison to
solar prominences, which are often quite flat. Here we obtain
$h_0=0.077~R_\odot$, relatively close to the estimated initial
prominence height of $\approx\!(0.05\mbox{--}0.06)~R_\odot$ (Paper~I).
If we would instead choose to compare the simulations to the temporal
profile of the prominence rotation, then each change of the twist,
which implies a change of the KI growth rate, would require a
rescaling of the time unit in the simulations, $\tau_A$. The
comparison of the simulated rotation profiles with the observed
profile is displayed in Figure~\ref{f:rotation}. As discussed in
Sections~\ref{s:intro} and \ref{s:discussion}, we disregard the slight
backward rotation at $h\gtrsim1.5~R_\odot$ above the photosphere in the
comparison and assume that the tendency of the rotation to level off at
this height would have continued in the absence of the specific
complex structure of the large-scale coronal field above AR~10989 and
in the absence of the heliospheric current sheet, which are not
included in our model. Several conclusions can be drawn from this set
of simulations.

(1) Similar height-rotation profiles (not only a similar total
rotation) are obtained in a range of $\Phi$-$B_\mathrm{et}$
combinations. The profiles for
$(\Phi,B_\mathrm{et}/B_\mathrm{ep})=(5\pi,0.42)$, $(3.5\pi,0.67)$, and
$(2.5\pi,1.06)$ all match the observed profile very well up to a
height $h\sim20~h_0\approx1.5~R_\odot$ above the photosphere, where a
total rotation of $\approx\!115\degrees$ is observed. These runs
include a strongly and a weakly kink-unstable and a kink-stable case.
Hence, even such a strong rotation does not by itself imply the
occurrence of the helical kink instability. Further arguments, such as
those given below, are required to draw conclusions about the
occurrence of the instability in the modeled event.

(2) To reach the observed total rotation of $\approx\!115\degrees$
with the initial configuration and parameter settings chosen in this
series, in particular with the chosen value of the sunspot semi-distance
$L$, the shear must contribute. The strongly twisted configuration
($\Phi=5\pi$) yields only little more than one third of the observed
rotation in the absence of shear ($B_\mathrm{et}=0$). Therefore, the
shear contributes the main part of the total rotation even in this
strongly kink-unstable case. Note that this conclusion changes if the
sunspot distance is set to larger (however, unrealistic) values, so
that the overlying field decreases less steeply with height (see
Section~\ref{ss:poloidal_field}).

(3) The twist also contributes in all runs. The tension of the twisted
field relaxes in any case when the flux rope is driven upward out of
its initial equilibrium, be it by the helical kink instability, by the
torus instability, or by any other process (\eg, by so-called
tether-cutting reconnection). This relaxation contributes to the
writhing of the flux rope axis regardless of whether or not the
helical kink instability is triggered. As a consequence, we do not
observe a jump in the achieved rotation as the twist of the initial
equilibrium is varied between kink-stable and kink-unstable values.
This is most obvious from the runs with $B_\mathrm{et}=0$.

(4) The higher the relative contribution of the twist, the lower the
height range where most of the rotation is reached. This reflects the
fact that the KI tends to reach saturation quickly, often already
when the flux rope has risen to a height comparable to the footpoint
distance (\eg, \opencite{torok:2004}). This property corresponds well
to the tendency of the rotation to level off at the relatively low
height of $\approx\!1.5~R_\odot$ ($\approx\!20~h_0$) above the photosphere.
The rotation by the shear field acts in a larger height range.
The different behavior can be made plausible from the fact that the
Lorentz force due to the shear field depends on the current through
the rope and on the angle between the flux rope legs and the shear
field. While the current decreases as the rope ascends (similar to the
twist), the angle rises until the legs approach a vertical position,
which corresponds to bigger apex heights than the saturation height of
the helical kink mode. Hence, the Lorentz force due to the shear field
acts strongly in a larger height range than the tension force
associated with the twist.

As a consequence, the Titov-D\'emoulin flux rope with
sub-critical twist for KI onset does not allow to match the entire
observed rotation profile of the 9~April 2008 event. We have performed
considerable numerical experimenting in this range of twists
[$\Phi=(2.5\mbox{--}3)\pi$], including modifications of the height
profiles $B_\mathrm{et}(z)$ and $B_\mathrm{ep}(z)$ and of the flux
rope shape (by varying its major radius $R$ but not the apex height
$h_0$) from the uniform settings for the runs in
Figure~\ref{f:rotation}. Either the rotation in the height range
$h\lesssim20~h_0$ was found to be too small, or the total rotation at
$h=30~h_0$ was too large. Although the shape of the prominence in the
plane of the sky can still be met by the special selection of the
field lines in Figure~\ref{f:2.5pi}, the saturation of the rotation at
$h\approx1.5~R_\odot$, revealed by the stereoscopic reconstruction,
cannot be reproduced. This suggests that at least a weak helical kink
instability must have been triggered in this event.

\begin{figure}
\centering
\includegraphics[width=0.9\textwidth]{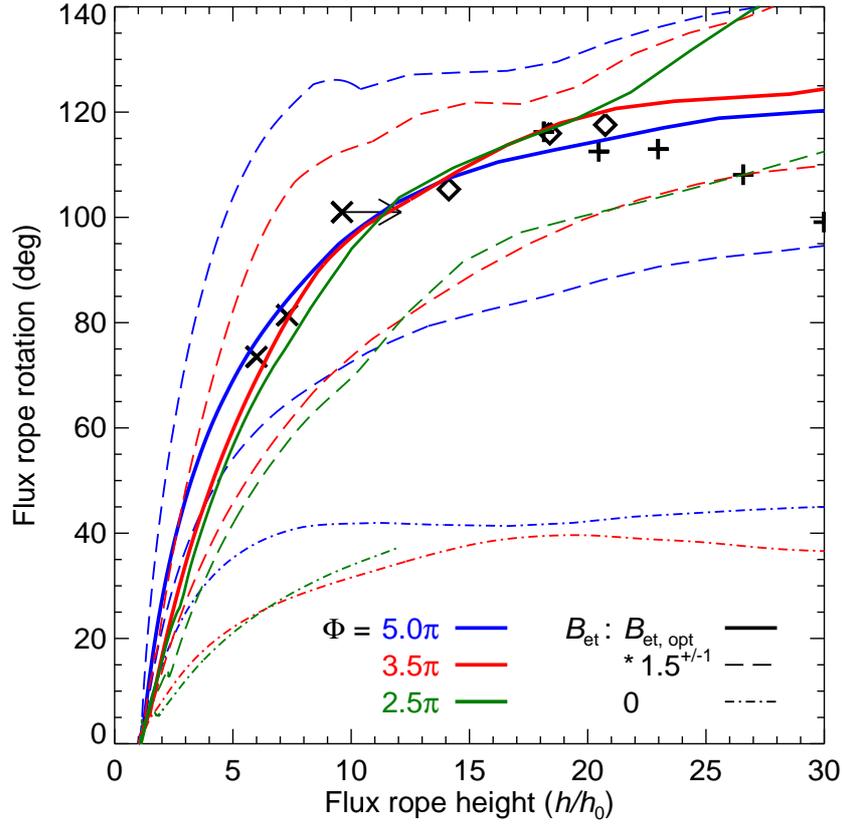}
\caption[]
{Comparison of flux rope rotation as a function of normalized apex
 height above the photosphere with the observation data obtained in
 Paper~I. Crosses and diamonds are EUVI data, with the final cross
 representing a lower limit for the height and the diamonds
 representing interpolated heights. Plus symbols are COR1 data.
 The distance between the footpoints of the flux rope in the
 simulation is scaled to the value of 175~Mm estimated in Paper~I,
 resulting in $h_0=0.077~R_\odot$.
 The initial average twist, $\Phi$, and the strength of the shear
 field component (external toroidal field), $B_\mathrm{et}$, given by
 its ratio to the external poloidal field component $B_\mathrm{ep}$ at
 the initial flux rope apex, are varied, while the geometrical
 parameters of the initial flux rope (except the minor radius $a$) and
 the spatial structure of the external field components
 $B_\mathrm{et}$ and $B_\mathrm{ep}$ are uniformly chosen throughout
 the series of runs (see Section~\ref{s:model} for detail). The
 optimum values for the shear field strength, which yield the best
 match with the observed rotation profile up to $h\approx20~h_0$, found
 through parametric search, are
 $B_\mathrm{et,\,opt}/B_\mathrm{ep}=0.42$, 0.67, and 1.06 for
 $\Phi=5.0\pi$, $3.5\pi$, and $2.5\pi$, respectively. Changes of
 $B_\mathrm{et}$ by a factor 3/2 and the case $B_\mathrm{et}=0$ are
 included.} 
\label{f:rotation}
\end{figure} 

(5) The range of twist-shear combinations that reproduce the observed
rotation profile is bounded not only from below, as outlined in (2)
and (4), but also from above. Average twists significantly exceeding
$5\pi$ are not only unlikely to occur in the corona but also lead to
increasingly strong helical deformations of the flux rope, which are
favorable for the onset of magnetic reconnection with the overlying
field or between the flux rope legs. Such reconnection can strongly
distort the rotation profile and can even stop the rise of the flux
rope \cite{torok:2005,Shiota&al2010}. Reconnection with the overlying
field does indeed lead to a confined (failed) eruption in the present
simulation series when the initial twist is raised to $6\pi$.
Reconnection between the legs of the rope occurs if $\Phi\ge7\pi$,
also leading to confined eruptions. (A detailed description of such
reconnection can be found in \opencite{Kliem&al2010}.)

Increasing the shear field tends to stabilize the flux rope because
any displacement then requires an increasing amount of energy to push
the ambient field aside. The low-twist case ($\Phi=2.5\pi$) with the
strongest shear field included in Figure~\ref{f:rotation} requires a
considerable initial perturbation to reach the torus-unstable range of
heights ($h>2.6~h_0=0.2~R_\odot$ for these parameters); it is completely
stable to small perturbations. Similarly, while the $3.5\pi$ run with
$B_\mathrm{et}=0$ is clearly kink-unstable, the corresponding sheared
case ($B_\mathrm{et}/B_\mathrm{ep}=0.67$) exceeds the instability
threshold only slightly. The initial lifting of the flux rope required
in the low-twist case strongly exceeds the observed rise of the
prominence to $\approx\!0.06~R_\odot$ prior to the onset of the eruption.
This represents a further strong indication against this
configuration.

The upper limit for the shear field is not a universal number but
depends on other parameters of the system, which include the thickness
of the flux rope, the strength of the line tying, and the height
profile of the external poloidal field, $B_\mathrm{ep}(z)$. A
systematic study of these dependencies would be beyond the scope of
the present investigation. However, we have considered a change of the
height profile $B_\mathrm{ep}(z)$, which is the key parameter for
the onset of the torus instability in the absence of shear and
significant line tying \cite{kliem:2006}. In an attempt to ease the
occurrence of the instability in the low-twist case ($\Phi=2.5\pi$,
$B_\mathrm{et}/B_\mathrm{ep}=1.06$), the sunspot semi-distance was reduced to
the minimum value of the possible range estimated from the observations,
$L=0.4$, leaving the other parameters of the equilibrium unchanged. No
reduction of the minimum height for instability was found, which must
be due to the strong stabilizing effect by the chosen shear field.

(6) Reconnection of the flux rope legs with the ambient field
contributes only a minor part of the total rotation in our simulation
series. It appears to remain weaker than the twist-driven rotation, or
at most comparable, \ie, considerably weaker than the shear-driven
rotation. This can be seen most clearly in the $5\pi$ run with
$B_\mathrm{et}=0$. Here the reconnection of the flux rope legs with
the ambient field proceeds while the rope apex rises from
$\approx\!2~h_0$ to $\approx\!16~h_0$, with the flux in the core of the
rope being involved in the range of apex heights
$h\sim(4\mbox{--}16)~h_0$. However, the major part of the total
rotation of $\approx\!40\degrees$ is already reached at low apex
heights, $h\lesssim5~h_0$, \ie, due to the helical kink mode. The apex
height range during the reconnection of the flux rope legs in the
shear-free $3.5\pi$ run is similar to the $5\pi$ run. The rotation
profile of this run in Figure~\ref{f:rotation} shows about equal
amounts of rotation in the height ranges $h\lesssim5~h_0$ and
$h\sim(5\mbox{--}16)~h_0$, indicating that the reconnection-driven
rotation could here be comparable to the twist-driven rotation. Again,
both remain considerably smaller than the rotation due to the shear in
the $3.5\pi$ run that best fits the observation data.

These conclusions are also supported by the fact that the angular
distance between the initial and new footpoints of the flux rope's
magnetic axis, measured from ${\bf x}=0$, remains far smaller than the
total rotation of the rope (see the right panels in
Figures~\ref{f:5pi}--\ref{f:2.5pi}).

\subsection{Influence of the External Poloidal Field}
\label{ss:poloidal_field}

\begin{figure}
\centering
\includegraphics[width=0.9\textwidth]{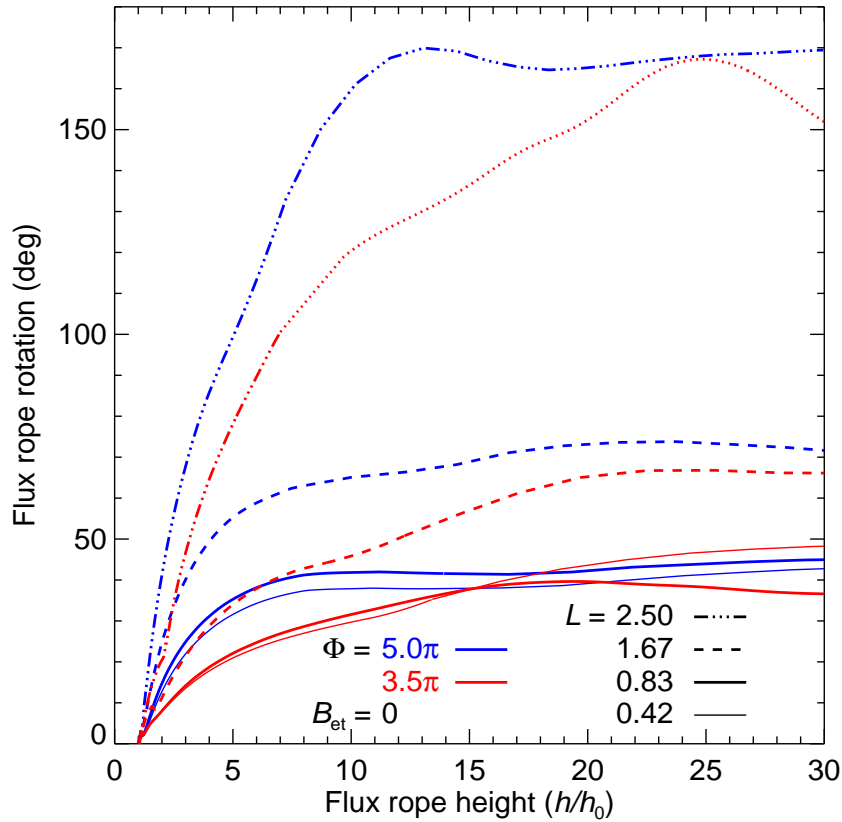}
\caption[]
{Dependence of flux rope rotation \vs\ height upon the distance $L$
 of the main flux concentrations in the source region from the PIL.
 Kink-unstable flux ropes ($\Phi=5\pi$ and $3.5\pi$) are considered
 for vanishing external shear field component, $B_\mathrm{et}=0$.}
\label{f:rotation_L}
\end{figure}

The height profile of the poloidal field which is due to sources
external to the flux rope, $B_\mathrm{ep}$, is a further factor of
potentially strong influence on the rotation. Erupting flux ropes
rotate more strongly at low heights if the external field initially
overlying the flux rope decreases more gradually with increasing
height \cite{Torok&al2010}. The relaxation of the magnetic tension in
the erupting flux rope by rotation is then more pronounced because the
relaxation by upward expansion is hindered, at least initially. The
relevant length scale,
$l_z=-[\mathrm{d}(\log B_\mathrm{ep})/\mathrm{d}z]^{-1}$, increases
with increasing distance between the sources of $B_\mathrm{ep}$, \ie,
between the main flux concentrations to the sides of the PIL. This can
easily be seen for the Titov-D\'emoulin equilibrium, where this scale
height is $l_z=(z+d)[1+L^2/(z+d)^2]/3$.

Figure~\ref{f:rotation_L} shows that this effect remains weak as long
as the distance between the sources of $B_\mathrm{ep}$, $2L$, is smaller than
distance between the footpoints of the erupting flux rope,
$2D_\mathrm{f}$, but that it
becomes very strong when the reverse relation holds. Here the sunspot
semi-distance $L$ is varied for the $5\pi$ and $3.5\pi$ runs with no
external shear field, $B_\mathrm{et}=0$, to be 0.5, 1, 2, and 3 times
the value estimated from the observations and used in
Section~\ref{ss:rotation_profile} 
(Figures~\ref{f:5pi}--\ref{f:rotation}). The two distances are nearly equal
if $L$ is set to twice the
estimated value. This is larger than the maximum of the range for $L$
compatible with the observations (see the Introduction). Hence, the
conclusions drawn from the series of simulations shown in
Figure~\ref{f:rotation} are not sensitive to the actual value of the
parameter $L$ as long as it remains within this range. In particular,
an external shear field component of strength close to the optimum
values given in this figure is then required to reach the observed
rotation.

Rotations even exceeding those produced mainly by the shear field in
Figure~\ref{f:rotation} are achieved in the absence of a shear field
for both twists if $L$ exceeds $D_\mathrm{f}$
by a factor $\gtrsim1.5$. A similar situation was
realized in simulations of erupting flux ropes in
\inlinecite{fan:2003} and \inlinecite{Gibson&Fan2008}, which showed
strong rotations of 115--120 degrees with $B_\mathrm{et}=0$. However,
such large distances of the main polarities, relative to the length of
the PIL and a filament channel between them, do not typically occur in
fully developed active regions. Hence, the effect of a shear field
\cite{isenberg:2007} will typically be involved if erupting flux
rotates by large angles of order 90\degrees\ and more.

\subsection{Rise Profile}
\label{ss:rise_profile}

The results of
Sections~\ref{ss:rotation_profile}--\ref{ss:poloidal_field} lead to
the question whether the initial twist and the shear field in the
source volume of the eruption can be further constrained individually,
although their combined effect on the rotation is similar. The
rotation profile obviously is a powerful new diagnostic of the
evolution of flux ropes in CMEs, however, for the considered event it
does not allow to discriminate between the strongly and weakly
kink-unstable cases shown in Figures~\ref{f:5pi} and \ref{f:3.5pi},
respectively. Therefore, we now consider the rise (time-height)
profile of the erupting flux. This function reflects the growth rate
of the instability driving the eruption. The growth rate varies
strongly with the twist if this parameter exceeds the threshold of the
helical kink mode (see, \eg, Figure~5 in \opencite{torok:2004}). When
the variation of the twist is combined with a variation of the shear
field strength in the opposite direction (one increasing, the other
decreasing), such that the rotation profile stays nearly unchanged,
then the rise profile will change even stronger: decreasing
(increasing) shear field strength leads to higher (lower) KI growth
rate. Thus, the combined comparison can constrain these parameters
individually.

In order to compare the simulated rise profiles with the observed one,
the time unit in the simulations, $\tau_A$, must also be scaled to a
dimensional value. Since $\tau_A=h_0/V_\mathrm{A0}$ and $h_0$ is
already scaled, this is equivalent to adopting a value for the initial
Alfv\'{e}n velocity $V_\mathrm{A0}$ in the body of the prominence. So
far, this parameter can hardly be derived from observations, since
both the field and density structure of prominences are generally only
poorly known. Therefore, here we work backwards by first finding the
best match between the simulated and observed rise profiles and then
checking whether the implied Alfv\'{e}n velocity falls within an
acceptable range. Lower bounds on the Alfv\'{e}n velocity in filaments
have been obtained through the application of seismological techniques
to six cases of oscillating filament threads \cite{Terradas&al2008}.
Five of these lie in the range $\sim\!(300\mbox{--}600)$~km\,s$^{-1}$
if the length of the field lines that pass through the threads is
assumed to be $\sim\!175$~Mm, the length of the erupting structure
estimated in Paper~I. An upper bound of order 1000~km\,s$^{-1}$ is
widely accepted for old, dispersed active regions like the one
considered here.

\begin{figure}
\centering
\includegraphics[width=0.9\textwidth]{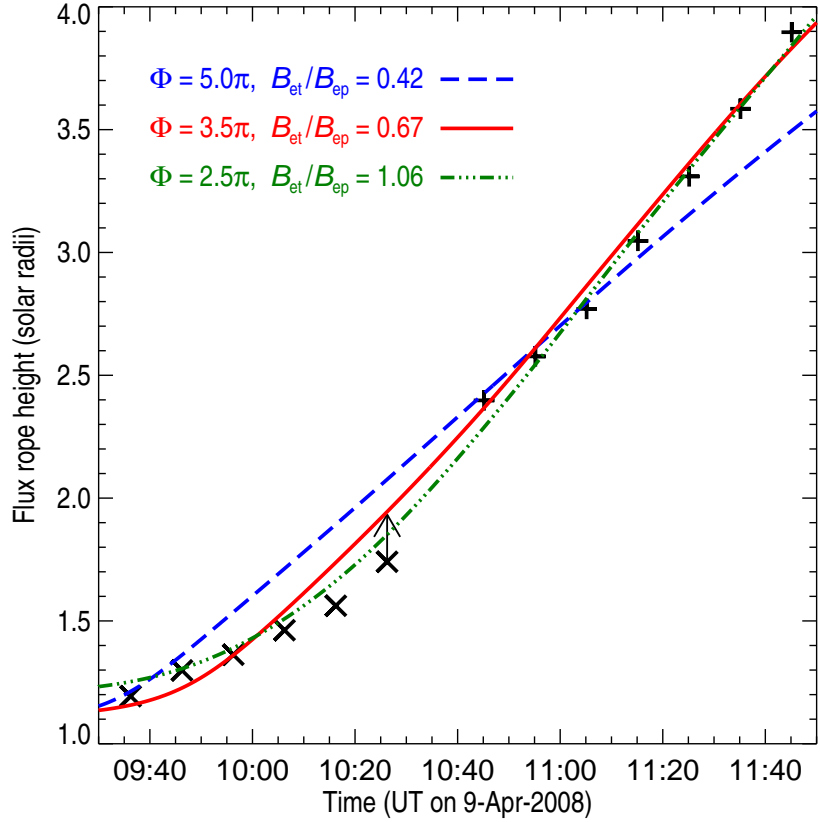}
\caption[]
{Comparison of the observed and simulated rise profiles of the flux
 rope apex, using the same scaling of lengths in the simulations as in
 Figure~\ref{f:rotation} and a start time of the eruption at
 08:48:00~UT. EUVI and COR1 data from Paper~I are plotted using the
 same symbols as in Figure~\ref{f:rotation}.
 The $5\pi$, $3.5\pi$, and $2.5\pi$ runs of
 Figures~\ref{f:5pi}--\ref{f:2.5pi} are scaled to these data assuming
 Alfv\'{e}n velocities $V_\mathrm{A0}=420$~km\,s$^{-1}$,
 550~km\,s$^{-1}$, and 560~km\,s$^{-1}$, respectively.}
\label{f:rise}
\end{figure}

The rise profiles of the simulation runs shown in
Figures~\ref{f:5pi}--\ref{f:2.5pi} are scaled and matched to the
observed profile in Figure~\ref{f:rise}. In selecting the scaling
parameters for the best match, we adopt a start time of the eruption a
couple of minutes before 08:51~UT, as estimated in Paper~I. The
conclusions drawn from the comparison do not depend upon the
particular start time if chosen in this range. The value 08:48~UT used
in Figure~\ref{f:rise} yields the best match of the $3.5\pi$ and
$2.5\pi$ runs with the observations and lies very close to (30~sec
before) the last EUVI image prior to the occurrence of motions in the
prominence along the path of the CME. Also, we give relatively low
priority to the EUVI height data after 10~UT, since these may be
smaller than the true heights, as discussed in Paper~I.

The scaled rise profile of the simulation with $\Phi=3.5\pi$ is found
to fit the data quite well if the Alfv\'{e}n velocity is chosen in the
range $V_\mathrm{A0}=(540\mbox{--}560)$~km\,s$^{-1}$ and the start
time of the simulation is placed in the range 08:45--08:50~UT (with
the earlier time corresponding to the lower $V_\mathrm{A0}$). These
values appear very plausible.

We did not succeed to find a satisfactory fit by the higher twisted
case. The corresponding curve in Figure~\ref{f:rise} demonstrates
this, using the same start time as for the $3.5\pi$ run and
$V_\mathrm{A0}=420$~km\,s$^{-1}$. Increasing (decreasing)
$V_\mathrm{A0}$ leads to a steeper (flatter) fit curve, \ie, to a
better fit at the larger (smaller) heights (if the start time is
adjusted simultaneously), but it is obvious that the curve can never
fit the combined EUVI and COR1 time-height data. Here the phase of
accelerated rise ends too early because the instability grows and
saturates too quickly. The rise profile of this simulation can be
stretched on the time axis and formally be fit to the data if in
addition to an unrealistically low Alfv{\'e}n velocity of
300~km\,s$^{-1}$ (lower than the terminal speed of the CME core) an
unrealistically large extension of the prominence flux of 360~Mm
(twice as large as the estimate in Paper~I) are assumed. Both are not
acceptable. This comparison with the data thus argues clearly against
the occurrence of high twist and a strong helical kink instability in
the considered event, in spite of the high total rotation.

Assuming the same start time as for the other two runs, the
kink-stable low-twist case ($\Phi=2.5\pi$) allows an acceptable
approximation of the observed rise profile, which yields a plausible
value of 560~km\,s$^{-1}$ for the Alfv{\'e}n velocity. The match is
slightly worse in comparison to the $3.5\pi$ run because the curve
does not reach the height of the first COR1 data point. Reducing
$V_\mathrm{A0}$, and adjusting the start time, allows for a nearly
perfect match of the COR1 data, similar to the $3.5\pi$ run, but this
moves the simulation curve, which already runs above all EUVI data
points, further away from the measurements in this height range, so
that the overall match is degraded.

The origin of the difference lies in the tendency of the torus
instability to spread the main upward acceleration of the flux across
a larger height range than the helical kink instability, which can be
clearly seen in Figure~\ref{f:rise}. The height range for the torus
instability is small only if the field in the source volume of the
eruption decreases very rapidly with distance from the flux rope
position (see Figure~1 in \opencite{kliem:2006}), \ie, in very compact
active regions of high field strength, especially in quadrupolar ones.
Since AR~10989 was already rather diffuse by the time of the eruption,
there is no justification to make the initial configuration in the
simulations more compact for a better fit of the rise profile by the
kink-stable configuration.

\subsection{Implications for the 9 April 2008 Eruption}
\label{ss:implications}

\begin{figure}
\centering
\includegraphics[width=0.9\textwidth]{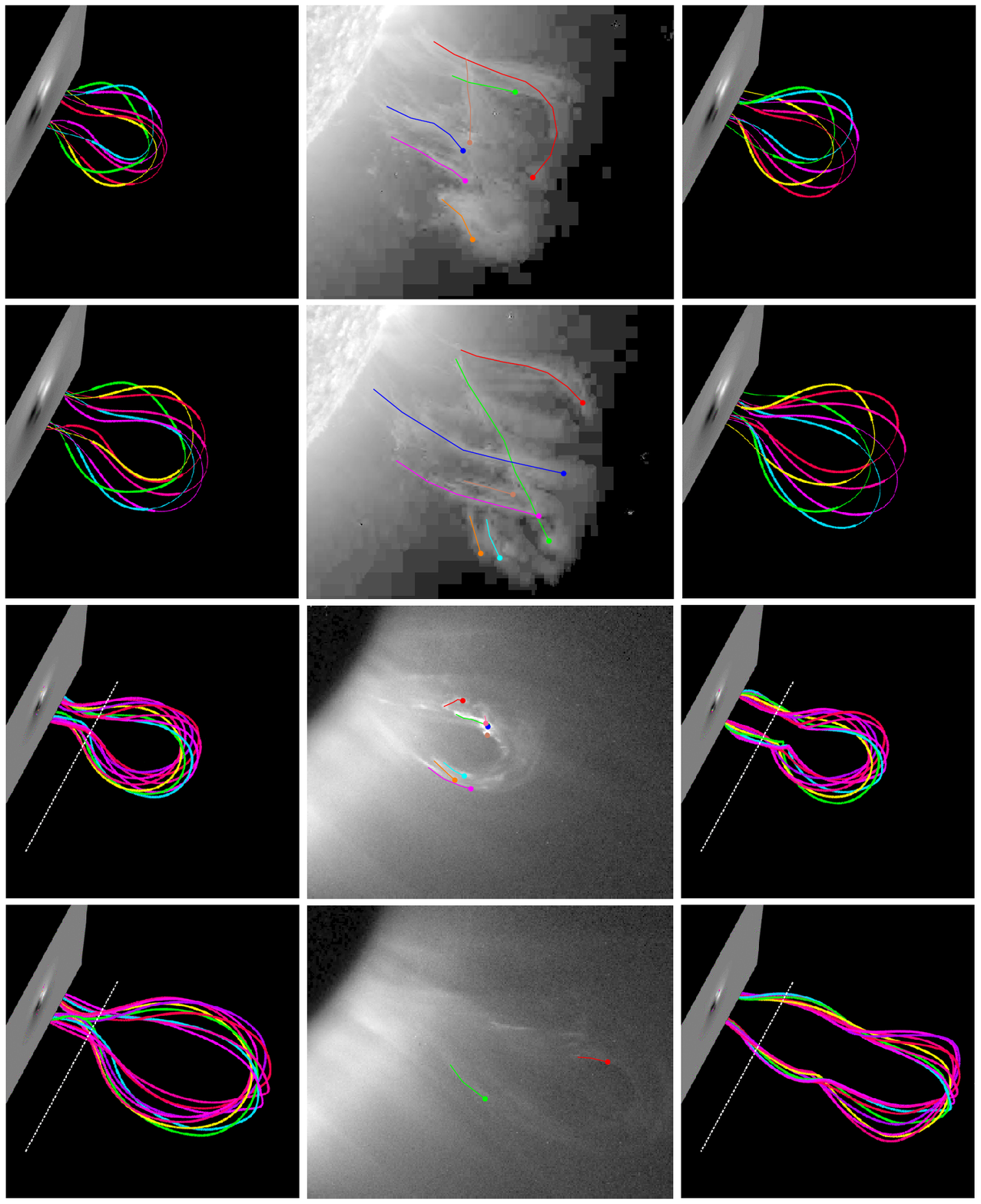}
\caption[]
{Comparison of simulated and observed flux rope shape for the
 kink-stable run ($\Phi=2.5\pi$, $B_\mathrm{et}/B_\mathrm{ep}=1.06$;
 left panels) and the weakly kink-unstable run ($\Phi=3.5\pi$,
 $B_\mathrm{et}/B_\mathrm{ep}=0.67$; right panels) in our parametric
 search which best match the observed rotation and rise profiles in
 their entirety. The STEREO images from Figure~\ref{f:obs} are
 supplemented by an additional image at 10:26~UT from Paper~I.
 For both runs, some experimenting with the field line
 selection was performed until also the observed shape was matched best.
 This yielded a flux bundle running slightly under the apex point of the
 rope's magnetic axis for the kink-stable run, as in
 Figure~\ref{f:2.5pi}, and a flux bundle enclosing the axis for the
 kink-unstable run.}
\label{f:comp}
\end{figure}

Based on the good quantitative agreement of the simulated rotation and
rise profiles with the observations,
Sections~\ref{ss:rotation_profile}--\ref{ss:rise_profile} yield the
following picture. The rotation profile in the height range
$h\lesssim20~h_0\approx1.5~R_\odot$ above the photosphere is well
matched by a strongly kink-unstable case ($\Phi=5\pi$), a weakly
kink-unstable case ($\Phi=3.5\pi$), and a kink-stable case
($\Phi=2.5\pi$) if a shear field of appropriate strength is included
in each of them. At greater heights,
$h\approx(20\mbox{--}30)~h_0\approx(1.5\mbox{--}2.3)~R_\odot$, the
comparison yields a clear indication against the kink-stable case,
which enters this range with an accelerated rotation, while the
observed rotation levels off. The kink-stable case also requires a
considerably stronger initial perturbation, lifting the flux rope apex
into the torus-unstable range of heights, \ie, to
$h\gtrsim2.6~h_0=0.2~R_\odot$, a value not supported by the
observations. In comparison, the accelerated rise of the kink-unstable
cases in our simulation series starts essentially from
$h_0=0.077~R_\odot$, relatively close to the observed onset height of
$(0.05\mbox{--}0.06)~R_\odot$. The shear field required by the
kink-stable case is comparable to the external poloidal field, 
$B_\mathrm{et}/B_\mathrm{ep}=1.06$. In a bipolar region, this
corresponds to a similar distance between the main polarities along
and across the PIL, which is not supported by AR~10989 as long as its
magnetic structure could be discerned in the approach to the limb (see
Figure~4 in Paper~I). The rise profile rules out the strongly
kink-unstable case and yields a further indication against the
kink-stable case, albeit only a weak one.  Both the observed shape of
the flux rope as a whole and the observed angles between individual
threads and the rope axis can be approximately reproduced by all three
model systems, but the overall match is best for the weakly
kink-unstable case (Figures~\ref{f:5pi}--\ref{f:2.5pi}). This is
substantiated by Figure~\ref{f:comp}, where we plot the sets of field
lines for this and for the kink-stable case which were found to match
the observations closest, out of many different sets that were
considered.

The shape of the erupting flux rope's magnetic axis in the considered
event is not sufficiently well defined by the observations to allow a
clear discrimination between the three considered cases based on this
property alone. Note that for other events it has proven to be
decisive. For example, the shape of the two erupting filaments modeled
in \inlinecite{torok:2005} could be matched only if an initial average
twist of $5\pi$ was assumed, not with a twist of $4\pi$.

Overall, we conclude that both strongly kink-unstable and kink-stable
configurations can be excluded with a high degree of certainty,
leaving a weakly kink-unstable initial configuration as the most
likely source of the Cartwheel event. This configuration allows to
reproduce the event with observationally supported values for several
key parameters (flux rope length, distance of the main flux
concentrations, initial orientation) and with plausible assumptions
for the magnetic structure (flux rope in a simple bipolar active
region) and for the remaining free parameters (twist and shear field
strength).

Regardless of how definite the rejection of the other two cases is
considered to be, the rotation of the erupting flux was primarily
caused by a shear field \cite{isenberg:2007}. Weaker contributions
came from the relaxation of twist (most likely by a weak helical kink
instability) and from reconnection with the ambient field.

\section{Discussion}
\label{s:discussion}

The major simplifying assumptions adopted for the modeling in this
paper include (1) the neglect of the initial mainly axial propagation
of the prominence, (2) the neglect of any asymmetry and complexity
introduced by the large-scale overlying field, and (3) the assumption
of a well defined, coherent flux rope (\ie, the Titov-D\'emoulin
model). We discuss these here to assess their potential influence on
the results.

While the initial propagation of the prominence introduced an
asymmetry and, therefore, definitely had the potential to produce some
rotation, we expect that it could not contribute strongly because the
propagation was approximately along the flux holding the prominence.
This does not principally change the magnetic configuration and the
Lorentz forces which dominate the acceleration of plasma in the
low-beta corona.

The effects belonging to category (2) are likely to be relevant
primarily at considerable heights. AR~10989 was a relatively isolated
region of simple, bipolar structure, and this holds also for its
dispersed phase as long as it could be followed in the approach to the
limb. The potential-field source-surface extrapolation of the
photospheric field in Paper~I shows that the large-scale coronal field
associated with the polar fields and the heliospheric current sheet
began to dominate already at heights $h\gtrsim0.3~R_\odot$ above the
photosphere, where the horizontal field direction nearly reversed.
The force by the field component along the line between
the flux rope legs pointed in the direction of a clockwise rotation
above this height, opposite to the force low in the corona. However,
the shear field above $\sim\!0.3~R_\odot$ was weaker than the shear field
in the core of the active region by more than an order of magnitude,
so that it could efficiently counteract the continuing, oppositely
directed force by the shear field at low heights, and the angular
momentum of the already rotating flux rope, only by acting across a
considerably larger height range. This is consistent with the fact
that the possible weak reverse rotation occurred only at 
$h>1.5~R_\odot$ above the photosphere. Thus, the rotation caused by the
shear field and twist inside the bipolar active region (at
$h<0.3~R_\odot$) must have been dominant factors for the rotation in
the height range up to $\sim\!1.5~R_\odot$ modeled here. We cannot
exclude that the saturation of the rotation would have occurred at a
greater height if the horizontal field had not changed its direction
above the active region, however, this weakens only one of the three
main arguments against the kink-stable configuration summarized in
Section~\ref{ss:implications}. The saturation of the rotation profile,
at a very similar height, was also seen in another erupting quiescent
filament (\opencite{Bemporad&al2011}; see their Figure~5).

The effect of the heliospheric current sheet is expected to become
important only at even larger heights. Otherwise, the rotation would
not have shown the saturation near $h\sim1.5~R_\odot$ and the possible
subsequent slight reverse rotation; rather the continuation of the
rotation to the value of $\approx150\degrees$ found at $13~R_\odot$
would have proceeded already in the COR1 height range.

The assumption that erupting flux in CMEs takes the structure of a flux
rope is strongly supported by all available observations. Quantitative
differences to our modeling must occur when initial flux ropes of
different structure are used. These are not likely to be substantial if
only details of the structure differ. The helical kink mode is known to
not overly depend on the details of the current channel's radial
structure. This can be seen, for example, from the similar instability
thresholds found in \inlinecite{Mikic&al1990},
\inlinecite{Baty&Heyvaerts1996}, \inlinecite{torok:2004}, and
\inlinecite{fan:2003} although flux ropes with and without a net current
and with straight and arched geometries were investigated. Flux ropes
with hollow current channels have recently been found to be
representative of filament channels which have undergone substantial
amounts of flux cancelation (\eg, \opencite{Su&al2011}). It is
conceivable that their less compact current distribution leads to
smaller rotations than the Titov-D\'emoulin equilibrium with the same
twist. This will be a subject of future study. On the other hand, we
believe that a strongly kink-unstable configuration of this type would
likely still not match the observed rise profile. The structure and
strength of the external poloidal and toroidal field components do not
depend upon the details of the flux rope structure, so that two
aguments against the kink-stable configuration, which are based on the
required initial lifting and on the ratio of $B_\mathrm{et}$ and
$B_\mathrm{ep}$, would likely still apply.

An overlying current sheet \cite{Birn&al2003} may be of stronger
influence, but we have argued above that this was not the case for the
considered event at the low coronal heights modeled in this paper.

The situation likely changes if the flux rope is far less coherent
than the Titov-D\'emoulin configuration \cite{Green&al2011},
especially if it is split \cite{bobra:2008}. The investigation
how such complex cases might change our conclusions must be left for
future work.

The comparison of the flux rope rotations found in this paper with the
rotation in the simulation of a breakout CME by
\inlinecite{lynch:2009} suggests a strong dependence upon the
existence of a flux rope at the onset of the eruption. In that
simulation, the inflating flux of a continuously sheared arcade did
not show any significant rotation up to a heliocentric height of
$\approx\!2~R_\odot$. Flare reconnection commenced at this point, which
progressively transformed the inner part of the arcade into a growing
flux rope. The flux rope immediately began to rotate. This process was
monitored until the core of the rope reached a heliocentric height of
$\approx\!3.5~R_\odot$. Throughout this range, the rope showed a linear
increase of its rotation angle with height, and the twist in the rope
stayed below the threshold of the helical kink mode. The addition of
poloidal flux by flare reconnection was largely complete in the middle
of the height interval. The rotation profile in this model differs
principally from the data presented here, even if only the height
range $>\!2~R_\odot$ is considered, where a flux rope did exist. This
suggests that the presence of a flux rope at the onset of the eruption
was a key feature of the Cartwheel event.

An interesting result of our parametric study is that the erupting
flux rope did always show some amount of rotation, even in the
shear-free, kink-stable case included in Figure~\ref{f:rotation}. We
expect this to be generally valid if coherent force-free flux ropes 
are considered as the initial condition, because such ropes always
possess twist. An untwisted flux tube, known as a Theta pinch,
requires a radial pressure gradient to attain equilibrium. This is not
available if the plasma beta is very small, as expected for the lower
coronal part of active regions. Whether the observations support the
occurrence of rotation in essentially all events does not yet seem to
be clear. For example, \inlinecite{Muglach&al2009} report that only
about 10 cases of unambiguous rotation in erupting filaments not very
far from Sun center could be identified in the EUV observations by the
EIT instrument \cite{delaboudiniere:1995} for the whole solar cycle
23. However, many cases of only moderate rotation may remain
undetected in such data, due to the projection in the plane of the
sky. \inlinecite{Yurchyshyn&al2009} report 101 partial and full halo
CMEs which show a very broad distribution of the difference between
the estimated initial and final orientations at distances up to
$30~R_\odot$; these angles do not show a clustering at zero degrees.
However, they represent the net effect of rotation in the corona and
in the inner solar wind where the heliospheric current sheet likely
dominates. If the fraction of non-rotating events is relatively small,
then a plausible explanation is that other processes counteract
the rotation by twist relaxation and the shear field in these cases,
for example reconnection with the ambient field. If the fraction is
large, then such nearly exact cancelation of rotations is unlikely to
be the primary explanation. The implication would then be that the
current distribution in the erupting field is often less compact or
less coherent than in the Titov-D\'emoulin flux rope, including the
possibility that a flux rope does not yet exist at the onset of the
eruption.

\section{Conclusions}
\label{s:conclusions}

Our parametric study of force-free flux ropes which erupt from simple
bipolar source regions with no overlying current sheet and rotate
about the direction of ascent yields the following conclusions.

1) Both the force by an external shear field component
$B_\mathrm{et}$ \cite{isenberg:2007} and the relaxation of twist
$\Phi$ (\eg, \opencite{Torok&al2010}), are potentially very
significant contributors to the rotation.

2) For parameters typical of CME source regions, in particular if the 
sources of the external stabilizing field (usually the main flux
concentrations next to the PIL) have a smaller distance than the
footpoints of the erupting flux,
the shear field yields the dominant contribution to the rotation for a
wide range of shear field strengths. The relaxation of twist remains
the weaker contributor under these conditions, even if it is
sufficiently high to trigger the helical kink instability. However,
since twist always exists in force-free flux ropes, it always causes
at least some rotation. Strong rotations ($\gtrsim90\degrees$) can be
produced by the twist alone, but only for considerably larger
distances between the sources of the external stabilizing field than
typically observed.

3) The rotation in low-beta plasma is not guided by the changing
orientation of the PIL with height. For the geometrical conditions
typical of CME source regions, it is opposite in direction (see the
Appendix).

4) For a given chirality of the configuration, the external shear
field and the twist cause flux rope rotation in the same direction,
which is clockwise for right-handed field and counterclockwise for
left-handed field if seen from above.

5) The two processes are related to each other when considered in
terms of magnetic helicity. Both convert initial twist helicity of the
flux rope into writhe helicity. The same total rotation, and rotation
profiles which are very similar in a substantial part of the total
height range of rotation, result in a range of $B_\mathrm{et}$--$\Phi$
combinations.

6) The rotation due to twist relaxation tends to act mainly low in the
corona, in a height range up to only a few times the distance between
the footpoints of the erupting flux. The rotation by the shear field
tends to be distributed across a larger height range.

7) The mere fact that erupting flux rotates does not by itself imply
that the helical kink instability occurred.

8) The relative contributions to the total rotation by the shear field
and by the twist can be disentangled by comparing both the observed
rotation and rise profiles with the corresponding curves from a model,
since these profiles possess a different dependence upon the
$B_\mathrm{et}$--$\Phi$ parameter combination. The resulting estimate
for the twist allows one to judge the occurrence of the helical kink
instability.

9) Magnetic reconnection contributes only weakly (much less than the
shear field) to the total rotation in the simple bipolar source
regions considered.

\medskip 
From the comparison with the simulation of rotating flux in
\inlinecite{lynch:2009} we conclude:

10) The rotation profile differs strongly between configurations with
and without a flux rope at the onset of the eruption.

\medskip 
The comparison with the stereoscopic observations and
three-dimensional reconstruction of the erupting prominence in the
9~April 2008 ``Cartwheel CME'' additionally shows the following.

11) The rotation profile obtained in Paper~I from the stereoscopic
reconstruction of STEREO data is equally well reproduced by our model
up to heights $\approx\!1.5~R_\odot$ above the photosphere for a range of
$\Phi$--$B_\mathrm{et}$ combinations which include a strongly
kink-unstable case ($\Phi=5\pi$, $B_\mathrm{et}/B_\mathrm{ep}=0.42$),
a weakly kink-unstable case ($\Phi=3.5\pi$,
$B_\mathrm{et}/B_\mathrm{ep}=0.67$), and a kink-stable case
($\Phi=2.5\pi$, $B_\mathrm{et}/B_\mathrm{ep}=1.06$). However, the
strongly kink-unstable configuration is ruled out by the simultaneous
consideration of the rise profile, and several features of the
kink-stable model argue strongly against this configuration. These are
the implied high value of the shear field, the rotation profile at
greater heights, and the unrealistic start height of the unstable rise
of $\approx\!0.2~R_\odot$. Hence, the occurrence of a weak helical kink
instability in the Cartwheel event is very likely.

\medskip 
Our results add to the complexity of the phenomenon of flux rope
rotation in eruptions which is already known from investigations that
focused on the influence of reconnection (\eg,
\opencite{Jacobs&al2009}; \opencite{Shiota&al2010};
\opencite{Cohen&al2010}; \opencite{Thompson2011};
\opencite{Lugaz&al2011}). An overall very
complicated dependence on several parameters and on the structure of
the ambient field is revealed. Thus, the quantitative prediction of
the rotation is a difficult task. The parametric study performed here
indicates for simple bipolar source regions that the strength of the
external shear field is the primary parameter determining the total
rotation. The twist and the height profile of the external poloidal
field are of relatively minor importance as long as they stay in the
typical ranges indicated by the observations. We did not yet study a
possible influence of the
height profile of the external shear field. The external shear field
of filament channels may be estimated to sufficient precision from a
simple linear force-free field extrapolation. It will be worth testing
whether numerical modeling starting from such fields, embedded
in current-free outer field, yields rotations
in agreement with observations of eruptions from bipolar
source regions.

Several investigations indicate that erupting flux ropes align with
the heliospheric current sheet in the course of their interplanetary
propagation (\eg, \opencite{Bothmer&Schwenn1998};
\opencite{Yurchyshyn2008}; Paper~I). This suggests that the coronal
rotation merely decides whether a parallel or an antiparallel
alignment will result at 1~AU. However, since complex physics is
involved and since rotations on the order of 90\degrees\ may not be
rare, the quantitative study of the effects that determine the
rotation in the corona remains of high scientific and practical
interest.

\section*{Appendix}

\begin{figure}[t]
\centering
\includegraphics[width=\textwidth,clip=]{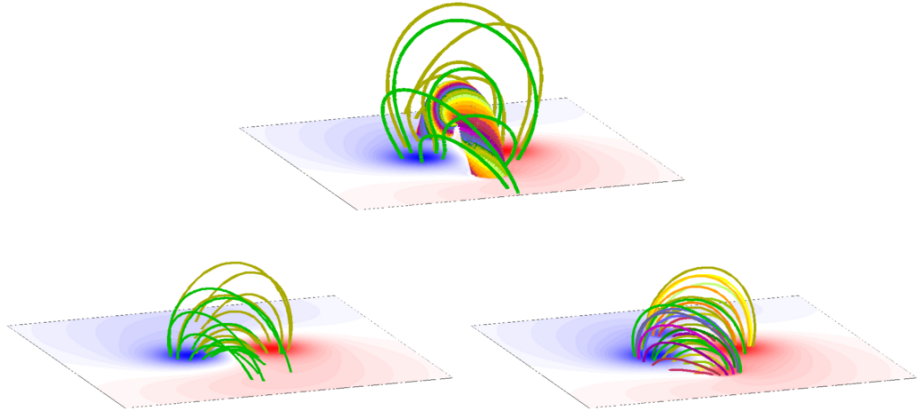}
\caption[]
{Visualization of the weakly kink-unstable modified Titov-D\'emoulin
 equilibrium ($\Phi=3.5\pi$, $B_\mathrm{et}/B_\mathrm{ep}=0.67$;
 Figures~\ref{f:TD99_3.5pi} and \ref{f:3.5pi}) whose eruption
 characteristics match the
 observations of the Cartwheel CME best (top panel) and of the
 corresponding external field (bottom left) and potential field
 (bottom right). The magnetogram and field lines starting in the
 photospheric flux concentrations are shown.}
\label{f:equilibrium}
\end{figure}

There are quite strong indications that CMEs align with the
heliospheric current sheet in the course of their propagation, \ie,
with the PIL in the solar wind (see references in
Section~\ref{s:conclusions}). This leads to the question whether the
PIL guides the rotation of erupting flux ropes also in the corona.
Here the PIL formed by the external field, due to sources outside the
flux rope, must be considered. We use ``CPIL'' to denote this
structure in the corona, where $\beta<1$. The heliospheric current
sheet and the CPIL differ in two properties of relevance here. First,
in the solar wind $\beta>1$, so that the pressure gradient is
generally dominant over the Lorentz force, while the opposite is true
in the corona. Second, the heliospheric current sheet is the location
of pressure gradients and Lorentz forces, while the CPIL generally
lacks both. In the low-beta corona, currents are induced at separatrix
surfaces, or at quasi separatrix layers, if the equilibrium is
perturbed or lost. The CPIL generally does not coincide with these
structures. Therefore, the CPIL should not influence the rotation of
erupting flux ropes in this height range.

\begin{figure}[!t]
\centering
\includegraphics[width=0.6\textwidth]{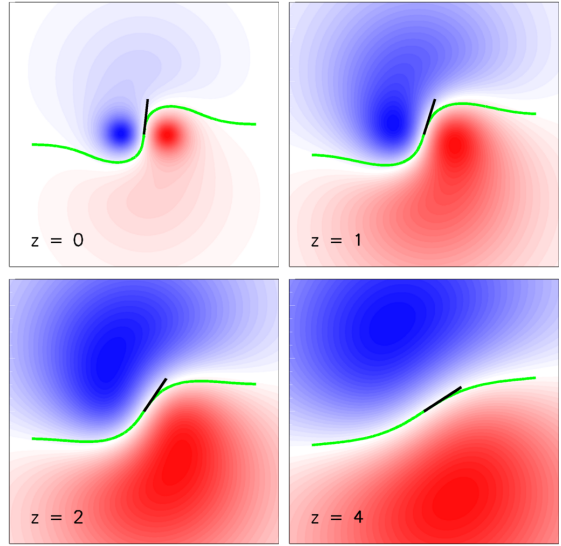}
\caption[]
{Orientation of the PIL in the external field of the configuration
 shown in Figure~\ref{f:equilibrium} at the position of the flux rope
 and different heights. The orientation is indicated by a black line.}
\label{f:CPIL}
\end{figure}

\begin{figure}[!ht]
\centering
\includegraphics[width=0.6\textwidth]{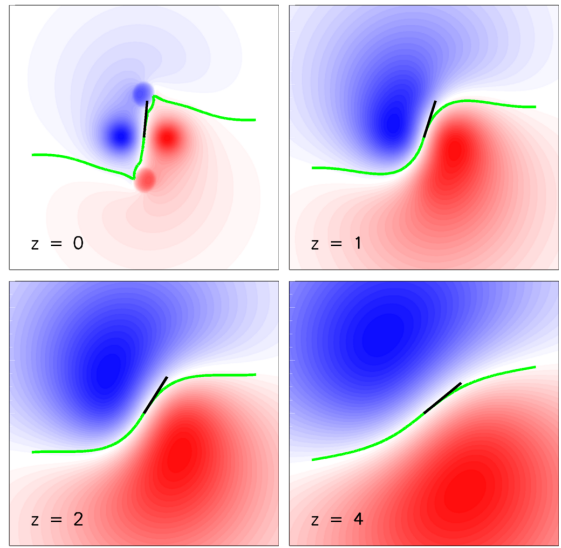}
\caption[]
{Same as Figure~\ref{f:CPIL} for the potential field of the
 configuration shown in Figure~\ref{f:equilibrium}.}
\label{f:BpCPIL}
\end{figure}

Figure~\ref{f:equilibrium} shows the initial equilibrium of the weakly
kink-unstable run which matches the Cartwheel event best, the
corresponding external field, and the potential field that results
when the full magnetogram of the vertical field component
of the equilibrium,
$B_z(x,y,0)$, is extrapolated into the volume above. The full
magnetogram includes the contributions from the flux
rope, which are excluded from the external field. The CPIL of this
configuration at the photospheric and three coronal levels is shown in
Figure~\ref{f:CPIL}. The CPIL changes its orientation in a clockwise
sense if one goes upward, but the unstable flux rope rotates in a
counterclockwise direction, since it is left handed. The clockwise
changing CPIL orientation results from the dominance of the external
toroidal field, $B_\mathrm{et}$, over the external poloidal field,
$B_\mathrm{ep}$, at great heights. This situation can typically be
expected to occur because $B_\mathrm{et}$ typically has a larger
spatial scale than $B_\mathrm{ep}$ (set by the distance between the
sources in the photosphere). The important fact here is that the CPIL
does not appear to have any significant influence on the rotation of
the flux rope in the zero-beta simulations performed in this paper.
For the reasons given above, this is valid also if other height
profiles of $B_\mathrm{et}$ or $B_\mathrm{ep}$ lead to a different
profile of the CPIL orientation with height.

Finally, we consider the approximation of the true CPIL by the PIL in
a potential-field extrapolation of the full photospheric magnetogram,
$B_z(x,y,0)$. In practice, it is difficult or even impossible to
determine the external field. This requires the determination of the
coronal currents through a nonlinear force-free extrapolation from a
vector magnetogram. The former is still difficult to carry out and the
latter may not be available. The PIL in the potential field
extrapolated from the magnetogram of the weakly kink-unstable
configuration in Figure~\ref{f:equilibrium} is shown in
Figure~\ref{f:BpCPIL}. Its orientation \vs\ height is very similar to
the behavior of the true CPIL. This supports the conclusions drawn in
Paper~I from a potential-field source-surface extrapolation for the
source region of the Cartwheel CME.

\begin{acks} 
We acknowledge the careful reading of the manuscript and the
constructive comments by the referee.
BK acknowledges support by the DFG, the STFC, and by NASA through 
Grant NNX08AG44G.
TT's work was partially supported by the European Commission
through the SOTERIA Network (EU FP7 Space Science Project No.\ 218816)
and by the NASA HTP and LWS programs.
WTT's work was supported by by NASA Grant NNG06EB68C.
\end{acks}

\bibliographystyle{spr-mp-sola-cnd} 
\bibliography{prominence_v4.2}

\end{article}
\end{document}